\documentclass[fleqn,usenatbib]{mnras}

\usepackage{newtxtext,newtxmath}

\usepackage[T1]{fontenc}

\DeclareRobustCommand{\VAN}[3]{#2}
\let\VANthebibliography\thebibliography
\def\thebibliography{\DeclareRobustCommand{\VAN}[3]{##3}\VANthebibliography}

\usepackage{graphicx}	
\usepackage{amsmath}	
\usepackage{subfig}
\usepackage[flushleft]{threeparttable}
\usepackage{booktabs}
\usepackage{anyfontsize}

\newcommand{\Rsun}{$R_{\odot}$}
\newcommand{\Rearth}{$R_{\oplus}$}

\newcommand{\tess}{\it TESS}

\title[Exploring the Neptunian Desert]{Exploring the Neptunian Desert: Insights from a Homogeneous Planetary Sample}

\author[L. Doyle et al.]{
Lauren Doyle,$^{1,2}$\thanks{E-mail: lauren.doyle@warwick.ac.uk}
David J. Armstrong,$^{1,2}$
Lorena Acu{\~n}a,$^{3}$
Ares Osborn,$^{4}$
S{\'e}rgio A. G. Sousa,$^{5,6}$
\newauthor
Amadeo Castro-Gonz{\'a}lez,$^{7}$
Vincent Bourrier,$^{8}$
Douglas Alves,$^{9}$
David Barrado,$^{7}$
Susana C. C. Barros,$^{5,6}$
\newauthor
Daniel Bayliss,$^{1,2}$
Kaiming Cui,$^{1,2}$
Olivier Demangeon,$^{5,6}$
Rodrigo F. D\'iaz,$^{10, 11}$
Xavier Dumusque,$^{8}$
\newauthor
Fintan Eeles-Nolle,$^{1,2}$
Samuel Gill,$^{1,2}$
Alejandro Hacker,$^{10}$
James S. Jenkins,$^{12,13}$
\newauthor
Marcelo Aron Fetzner Keniger,$^{1,2}$
Marina Lafarga,$^{1,2}$
Jorge Lillo-Box,$^{7}$
Isobel Lockley,$^{1,2}$
Louise D. Nielsen,$^{14}$
\newauthor
L{\'e}na Parc,$^{7}$
Jos{\'e} Rodrigues,$^{5,6,15}$
Alexandre Santerne,$^{16}$
Nuno C. Santos,$^{5,6}$
and Peter J. Wheatley$^{1,2}$
\\
$^{1}$Centre for Exoplanets and Habitability, University of Warwick, Coventry, CV4 7AL, UK \\
$^{2}$Department of Physics, University of Warwick, Coventry, CV4 7AL, UK \\
$^{3}$Max-Planck Institut für Astronomie, Königstuhl 17, 69117 Heidelberg, Germany \\
$^{4}$McMaster University, Department of Physics \& Astronomy, 1280 Main St W, Hamilton, ON L8S 4L8, Canada \\
$^{5}$Instituto de Astrof\'isica e Ci\^encias do Espa\c{c}o, Universidade do Porto, CAUP, Rua das Estrelas, PT4150-762 Porto, Portugal \\
$^{6}$Departamento\,de\,Fisica\,e\,Astronomia,\,Faculdade\,de\,Ciencias,\,Universidade\,do\,Porto,\,Rua\,Campo\,Alegre,\,4169-007\,Porto,\,Portugal \\
$^{7}$Centro de Astrobiolog\'ia (CAB, CSIC-INTA), Depto. de Astrof\'isica, ESAC campus, 28692, Villanueva de la Ca\~nada (Madrid), Spain \\
$^{8}$Observatoire Astronomique de l'Universit\'e de Gen\`eve, Chemin Pegasi 51, CH-1290 Versoix, Switzerland \\
$^{9}$Departamento de Astronom{\'i}a, Universidad de Chile, Camino el Observatorio 1515, Las Condes, Santiago, Chile \\
$^{10}$International Center for Advanced Studies (ICAS) and ICIFI (CONICET), ECyT-UNSAM, Campus Miguelete, 25 de Mayo y Francia, (1650) Buenos Aires, Argentina \\
$^{11}$ Instituto Tecnol\'ogico de Buenos Aires (ITBA), Iguaz\'u 341, Buenos Aires, CABA C1437, Argentina \\
$^{12}$Instituto de Estudios Astrof{\'i}sicos, Facultad de Ingenier{\'i}a y Ciencias, Universidad Diego Portales, Av. Ej{\'e}rcito 441, Santiago, Chile \\
$^{13}$Centro de Excelencia en Astrof{\'i}sica y Tecnolog{\'i}as Afines (CATA), Casilla 36-D, Santiago, Chile \\
$^{14}$University Observatory, Faculty of Physics, Ludwig-Maximilians-Universit{\"a}t M{\"u}nchen, Scheinerstr. 1, 81679 Munich, Germany \\
$^{15}$Observatoire François-Xavier Bagnoud -- OFXB, 3961 Saint-Luc, Switzerland \\
$^{16}$Aix Marseille Univ, CNRS, CNES, Institut Origines, LAM, Marseille, France
}
\date{Accepted XXX. Received YYY; in original form ZZZ}

\pubyear{2024}

\begin{document}
\label{firstpage}
\pagerange{\pageref{firstpage}--\pageref{lastpage}}
\maketitle

\begin{abstract}
In this paper, we present a homogeneous analysis of close-in Neptune planets. To do this, we compile a sample of TESS-observed planets using a ranking criterion which takes into account the planet's period, radius, and the visual magnitude of its host star. We use archival and new HARPS data to ensure every target in this sample has precise radial velocities. This yields a total of 64 targets, 46 of which are confirmed planets and 18 of which show no significant radial velocity signal. We explore the mass-radius distribution, planetary density, stellar host metallicity, and stellar and planetary companions of our targets. We find 26\% of our sample are in multi-planet systems, which are typically seen for planets located near the lower edge of the Neptunian desert. We define a `gold' subset of our sample consisting of 33 confirmed planets with planetary radii between 2$R_{\oplus}$ and 10$R_{\oplus}$. With these targets, we calculate envelope mass fractions (EMF) using the GAS gianT modeL for Interiors (GASTLI). We find a clear split in EMF between planets with equilibrium temperatures below and above 1300~K, equivalent to an orbital period of $\sim$3.5~days. Below this period, EMFs are consistent with zero, while above they typically range from 20$\%$ to 40$\%$, scaling linearly with the planetary mass. The orbital period separating these two populations coincides with the transition between the Neptunian desert and the recently identified Neptunian ridge, further suggesting that different formation and/or evolution mechanisms are at play for Neptune planets across different close-in orbital regions. 
\end{abstract}




\section{Introduction}
During the past three decades, over 5,800 {\footnote{\url{https://www.exoplanetarchive.ipac.caltech.edu/}}} exoplanets have been discovered, with new detections remaining a key objective in exoplanet science. Through ground-based and space-based observations, the properties of these exoplanets have been uncovered, with $\sim$70\% of exoplanets orbiting close to their stars with orbital periods of less than 30~days. This is partly due to an observational bias; short period planets are relatively easy to detect \citep{winn2015occurrence} and have a higher probability of being in a transiting configuration. However, their short orbital periods make them interesting targets as the increased stellar irradiation causes extreme atmospheric conditions. A planet's proximity to its star is directly linked to its surface temperature where for planets with the same atmospheric composition a shorter orbital period will increase the surface temperature. Therefore, we define planets as `hot' \citep[i.e. with P$_{\rm{orb}}$ < 10~days, see][]{dawson2018origins} or `warm' \citep[i.e. P$_{\rm{orb}}$ between 10 - 200~days, see][]{dawson2018origins, jackson2021observable, dong2021warm}. Planets with radii below $\sim$2~R$_{\oplus}$ typically have rocky structures and are commonly referred to as `Earths' and `super-Earths'. Large planets with radii above $\sim$10~R$_{\oplus}$ have extensive hydrogen and helium envelopes and are commonly referred as `Jupiters'. Planets in between are also known to typically have volatile-rich compositions (a mix of H/He with water, methane and other gaseous species), and are commonly referred to as `sub-Neptunes', `super-Neptunes', and `sub-Jovians/sub-Saturns', depending on their radii.

In the early era of exoplanet science, a correlation between mass and period of close-in Neptune-sized planets was suggested by \citet{mazeh2005intriguing}. This was strengthened by the increasing exoplanet population, where a significant dearth of Neptune-mass planets with orbital periods shorter than 2 -- 4~days was found \citep[e.g.][]{des2007diagram, davis2009evidence, szabo2011short, benitez2011mass, beauge2012emerging, helled2015possible}. This dearth cannot be explained by observational biases as Neptune-mass and Neptune-radius planets have been found on longer orbital periods \citep[see][]{howard2011nasa, orosz2012neptune, dong2013fast, courcol2015sophie}, and planets with large radii and short orbital period are easier to detect by radial velocities and transits \citep[see][and references therein]{winn2018kepler}. 

\citet{szabo2011short} completed the first cluster analysis in mass-period and radius-period space for all known exoplanets (106 at the time with precise masses, radii and orbital periods) to investigate the lack of hot Neptunes with orbital periods < 2.5~days. Overall, they found two distinct clusters amongst the exoplanet population, (i) Jupiters with P$_{\rm{orb}}$ between 0.8 -- 114~days with a narrow range of planet radii at 1.2 $\pm$ 0.2 R$_{\rm{J}}$ and (ii) a mixture of super Earths, Neptunes and hot Jupiters distributed tightly around orbital periods of 3.7 $\pm$ 0.8~days. From this initial analysis, the region of space known as the `Neptunian desert' was first established, and it has become one of the most prominent features amongst close-in exoplanets. The exact location, shape, and boundaries of the Neptunian desert have been mapped out over the years. Following a similar analysis as \citet{szabo2011short} but with an exoplanet sample size ten times larger, \citet{mazeh2016dearth} used the observed planet distribution to provide the first delimitation for the Neptunian desert in the period-radius and period-mass spaces. This desert is defined by an upper and a lower boundary intersecting at $\sim$10 -- 15~days, obtained through the identification of density contrasts in the distribution of hot Jupiters and small planets. Overall, \citet{mazeh2016dearth} speculated that the creation of the Neptune desert lies in the dynamical histories of giant planets. In recent years, Neptune desert planets have been increasingly observed, which led to several authors to propose different boundaries in the period-radius and insolation-radius spaces based on more recent distributions \citep[e.g.][]{2023A&A...677A..12D,2023A&A...671A.132S,magliano2024revisiting,2024A&A...690A..62P}. 

Recently, \citet{castro2024mapping} carried out a planet occurrence study to delineate the Neptunian desert and explore its transition into the `Neptunian savanna' \citep[i.e. a milder deficit of Neptune-size planets at longer periods;][]{bourrier2023dream} based on the underlying planet distribution. In the shortest-period orbits, the authors find a lower boundary with positive slope and an upper boundary with negative slope, similarly to \citet{mazeh2016dearth}. However, the bias-corrected boundaries do not intersect at $\sim$10 -- 15~days, but instead are limited by an over-density of planets in the $\sim$3.2 -- 5.7~days orbital period range, which the authors refer to as the `Neptunian ridge'. The authors interpret the ridge as an extension of the well-known hot Jupiter pileup \citep[][]{2003A&A...407..369U,2007ARA&A..45..397U} towards the Neptunian domain, which further suggests a key role of dynamical processes in shaping the desert \citep{dawson2018origins}.

There are two key formation models which lead to the creation of giant planets, core accretion \citep{pollack1996formation} and disk instability \citep{boss1997giant}. In core accretion, planet formation begins with the growth of planetesimals from dust which eventually forms into a solid core. Once solid material has been depleted the core continues to accrete gas forming an envelope, and runaway gas accretion begins if the initial core is large enough ($\sim$10 -- 20~M$_{\rm{\oplus}}$), creating a gas giant with a large envelope. For the disk instability model, there is gravitational fragmentation in the disk surrounding a young star, and clumps of gas and dust contract to form planets. Both of these models require the planet to form at large distances from the star. Occurrence rates of Neptunes and Jupiters are comparable over a wide range of orbital periods, implying a common formation pathway \citep{howard2012planet, bennett2021no}. The cores of Neptunes are also large enough for runaway accretion to occur, so some mechanism must have prevented them from becoming Jupiters. This could have been truncation of gas accretion through late formation of the planetary core \citep{batygin2016situ}, or early dissipation of the gas disk \citep{mordasini2011harps}. 

Other mechanisms can also be at play to alter the orbital and physical properties of close-in Neptunes. Giant planets can migrate inwards, shrinking their orbital periods. They could undergo several different migration processes. The first is disk driven migration, which would occur soon after their formation when the protoplanetary disk still exists  \citep{goldreich1979excitation, lin1996orbital, baruteau2016formation}. They could also undergo high eccentricity migration, caused by a massive outer companion, and this could happen at any stage in their lifetime \citep{wu2003planet, ford2008origins, chatterjee2008dynamical, correia2011tidal, beauge2012multiple}. Once a Neptune planet has reached a close-in orbit, evaporation can take over to cause mass loss via stripping of their gas envelopes \citep[e.g. see][]{ehrenreich2015giant, bourrier2018hubble, ben2022signatures, vissapragada2022upper}. It is the interplay between migration and evaporation which is believed to be responsible for shaping the landscape of Neptune planets we observe today \citep[see][]{bourrier2018orbital, owen2018photoevaporation}. 

In this paper, we look at the demographics of close-in Neptunian planets in as homogeneous a manner as possible. To do this, we select a sample of 64 {\tess} targets which lie in the original desert boundaries defined by \cite{mazeh2016dearth}, according to a ranking criterion which we detail in \S \ref{sec:sample}. In \S \ref{sec:multiplicity}, we explore the multiplicity of our sample, looking at both stellar and planetary companions within the systems. We then go on to discuss the sample in the context of the mass-radius plane in \S \ref{sec:density}, which allows for assumptions to be made on the planetary composition. We then explore planetary composition and interior structures further in \S \ref{sec:emf}, where we determine envelope mass fractions for a subset of our sample. We also explore the metallicity of our host target stars in \S \ref{sec:metallicity} and how this correlates with ultra-short period terrestrial and hot Jupiter planets. Finally, we discuss our findings and summarise our results in \S \ref{sec:discussion}.

\section{Sample Selection}
\label{sec:sample}
Investigating a planetary population in a homogeneous way requires a curated sample that allows acknowledgement of intrinsic biases. With this in mind, we choose to base our sample on the list of released {\tess} Objects of Interest (TOIs), which are candidate signals from the {\tess} mission, due to the sample size and single instrument source. 

Launched in April 2018, The Transiting Exoplanet Survey Satellite \citep[{\tess}:][]{ricker2015tess} has been crucial in increasing the number of transiting exoplanets found. Its core mission is to obtain photometric light curves of bright stars in order to search for potential transiting planets. The TOI list is created by searching for periodic flux decreases caused by potential transiting planets in both the Science Processing Operations Centre \citep[SPOC:][]{jenkins2016tessspoc, caldwell2020spoctargets} 2-minute light curves and 10 to 30 minute cadence Full Frame Images (FFI). These are then examined by the {\tess} Science Office (TSO) to identify planet candidates which would benefit from follow-up observations and released as TOIs \citep{guerrero2021tois}. 

The TOI list has inherent biases in the selection of stars for short cadence observations, as well as in the overall vetting process. Ideally, we would perform our own transit search and analysis \citep[see e.g.][]{2020AJ....159..248K,bryant2023occurrence} with an automated vetting process. At the time of observations of spectroscopic data used here, such a sample was not available and the TOI list represented the largest, most homogeneous set of close-in Neptunes. The TOI list used was generated in July 2022, marking the beginning of the radial velocity programme used in this study. 

\subsection{Radial Velocity Observations}
The Nomads programme (ESO program ID 108.21YY, P.I. Armstrong) aimed to measure the mass of $\sim$30 planets in order to substantially increase the number of Neptune desert planets with precise radii and mass measurements. The programme utilised the High Accuracy Radial velocity Planet Searcher \citep[HARPS;][]{pepe2002harps}, a cross-dispersed echelle spectrograph located at La Silla Observatory in Chile. It is mounted on the Cassegrain focus of the 3.6~m telescope, and operated by the European Southern Observatory (ESO). HARPS is capable of achieving a radial velocity (RV) precision of 1~ms$^{-1}$ with resolving power $R = 115,000$ and wavelength range between 390 -- 690~$nm$. 

In selecting targets for the programme, we followed the sample selection procedure below. Where targets had already been observed by HARPS or comparable spectrographs, we did not re-observe them, instead filling in `gaps' in the ranked list of targets such that the sample was complete down to a given merit function. Sample vetos and the merit function are described below. Most observed targets have been published separately by our programme and others (see Table~\ref{tab:planet_prop}). Observed targets which only provided an upper limit on planetary mass are published for the first time here, detailed in Appendix~\ref{sec:null_results}. In choosing this strategy, we prioritised defining and observing a complete sample, over and above homogeneity in observing \textit{strategy}, such as observing cadence and point number which varies target-to-target. Given that we are focusing on the properties of confirmed planets where the planetary signal has been significantly detected, and show the upper limits of cases where the planet could not be detected, we believe this strategy provides the most useful information available given limited spectroscopic observation time.

\subsection{Cuts on Desert Boundaries and Observability}

We build our sample based on the broad Neptunian desert boundaries defined by \citet{mazeh2016dearth}, as updated bias-corrected boundaries \citep[e.g.][]{castro2024mapping} were not available at the time. In order to identify TOIs meeting this criterion, we place limits on orbital period and planetary radius. We also limit the sample to stars suitable for precise RV characterisation, and visible with HARPS, making the following cuts:
\begin{enumerate}
    \item V$_{\rm{mag}}$ < 13: bright stars require less exposure time to reach an acceptable signal-to-noise and RV precision. 
    \item declination ($\delta) \leq$ +20$^{\circ}$: HARPS is located in the southern hemisphere; therefore, targets in the north of this declination are only observable over a short period of time.
    \item R$_{*} \leq$ 1.5~\Rsun: we remove larger A-type stars where the semi-amplitude of a Neptunian-sized planet would be too small to detect with HARPS. Targets with unknown stellar radii are also removed here, specifically TOI-161.
    \item T$_{\rm{eff}} \leq$ 6400~K: we remove hotter stars, as radius scales with effective temperature and hot stars rotate faster, broadening spectral lines. Additionally, hot stars are not ideal targets for precision RVs due to the reduced number of spectral lines with respect to cooler stars.
    \item Planets with orbital periods (P$_{\rm{orb}}$) and radii (R$_{p}$) must be inside the \citet{mazeh2016dearth} boundaries for the Neptunian desert. 
    \item P$_{\rm{orb}} \leq$ 10~days: cuts off the very tip of the \citet{mazeh2016dearth} desert triangle to focus on close-in planets.
    \item {\tess} Follow-up Observing Program (TFOP) Working Group (WG) Sub Group SG1\footnote{SG1 Observation Coordinator spreadsheet is not publicly available, applications for membership can be found at \url{https://tess.mit.edu/followup/apply-join-tfop/}}, focusing on photometry, planet candidate dispositions explicitly marked as `FA' or `EB' by the {\tess} TFOP are removed: this removes known False Alarms (FA) and Eclipsing Binaries (EB) which are not planets.
    \item We require at least one TFOP observation capable of resolving nearby eclipsing binaries, increasing the chance that the candidate is on-target.
\end{enumerate}

This process leaves a total of 192 TOIs which form a preliminary sample of close-in Neptune targets for follow-up with HARPS. During observations we also removed three individual targets from our list as a result of them being false positives which were found during our HARPS observations. These include: (i) TOI-2224 and TOI-2539 which are double-lined spectroscopic binaries, and (ii) TOI-835 which has strong stellar activity at double the candidate orbital period. Five further targets, including 2 upper limits, changed their nominal planetary radius during observations and fell outside the \citet{mazeh2016dearth} boundaries; this is the result of updated planetary radii values arriving as new {\tess} data was taken. These targets were removed as they retrospectively failed the above vetos.

\subsection{Merit Ranking}
\label{sec:merit}
We rank all 192 TOIs on a `merit function' which orders the targets based on their visual magnitude and position within the Neptune desert boundaries. As a result, our `merit function' ($M$) is a multiplication of two metric values, one corresponding to stellar magnitude ($M_{\rm{mag}}$), and the other for the planet's location within the desert boundaries in period-radius space ($M_{\rm{loc}}$). For stellar magnitude, the metric value is expressed as:

\begin{equation}
    M_{\rm{mag}} = |V_{\rm{mag}} - 13| + 1 .
\end{equation}

This scales linearly with magnitude where the smallest value is unity, having removed all targets with $V_{\rm{mag}}$ > 13. In terms of location within the desert boundaries, we use the definition of the boundaries from \citet{mazeh2016dearth} in period-radius space. We then modify these slightly to transform the boundary out of log space and add a minimum clause to the upper boundary. This is because at very short periods, the \citet{mazeh2016dearth} boundary extends to very large radii which if left would heavily bias the distance metric to ultra-short period planets. Therefore, adding the minima means the boundary does not extend to radii R$_{\rm{p}}$ > 10~\Rearth. The modified Neptunian desert boundaries are then expressed as:

\begin{equation}
\label{eq:ldb}
    R_{\rm{ldb}} = 10^{0.67\log_{10}(P_{\rm{orb}}) - 0.01},
\end{equation}

\begin{equation}
\label{eq:udb}
    R_{\rm{udb}} = min\{10, 10^{-0.33\log_{10}(P_{\rm{orb}}) + 1.7}\},
\end{equation}
where ldb stands for lower desert boundary and udb upper desert boundary. The metric based on location within the desert, $M_{\rm{loc}}$ is then expressed as the minimum of either equation \ref{eq:ldb} or \ref{eq:udb} as:

\begin{equation}
    M_{\rm{loc}} = min\{|R_{\rm{p}} - R_{\rm{ldb}}|, |R_{\rm{p}} - R_{\rm{udb}}|\}.
\end{equation}

The final merit function is expressed as a multiplication of the two metric values, $M = M_{\rm{mag}} \times M_{\rm{loc}}$. Since brighter targets require a shorter exposure time, resulting in less observing time being used, these are prioritised in our merit function. The magnitude of a star links to its spectral type, with increasing magnitude corresponding to fainter, lower mass main sequence stars \citep{shapley1921relation}, and, as with any magnitude-limited survey, we will have some bias towards hotter stars in our sample. Furthermore, we prioritise targets deeper within the desert due to their high importance when it comes to exploring the formation and evolution mechanisms behind sculpting the desert. 

\begin{figure}
    \centering
    \includegraphics[width=0.47\textwidth]{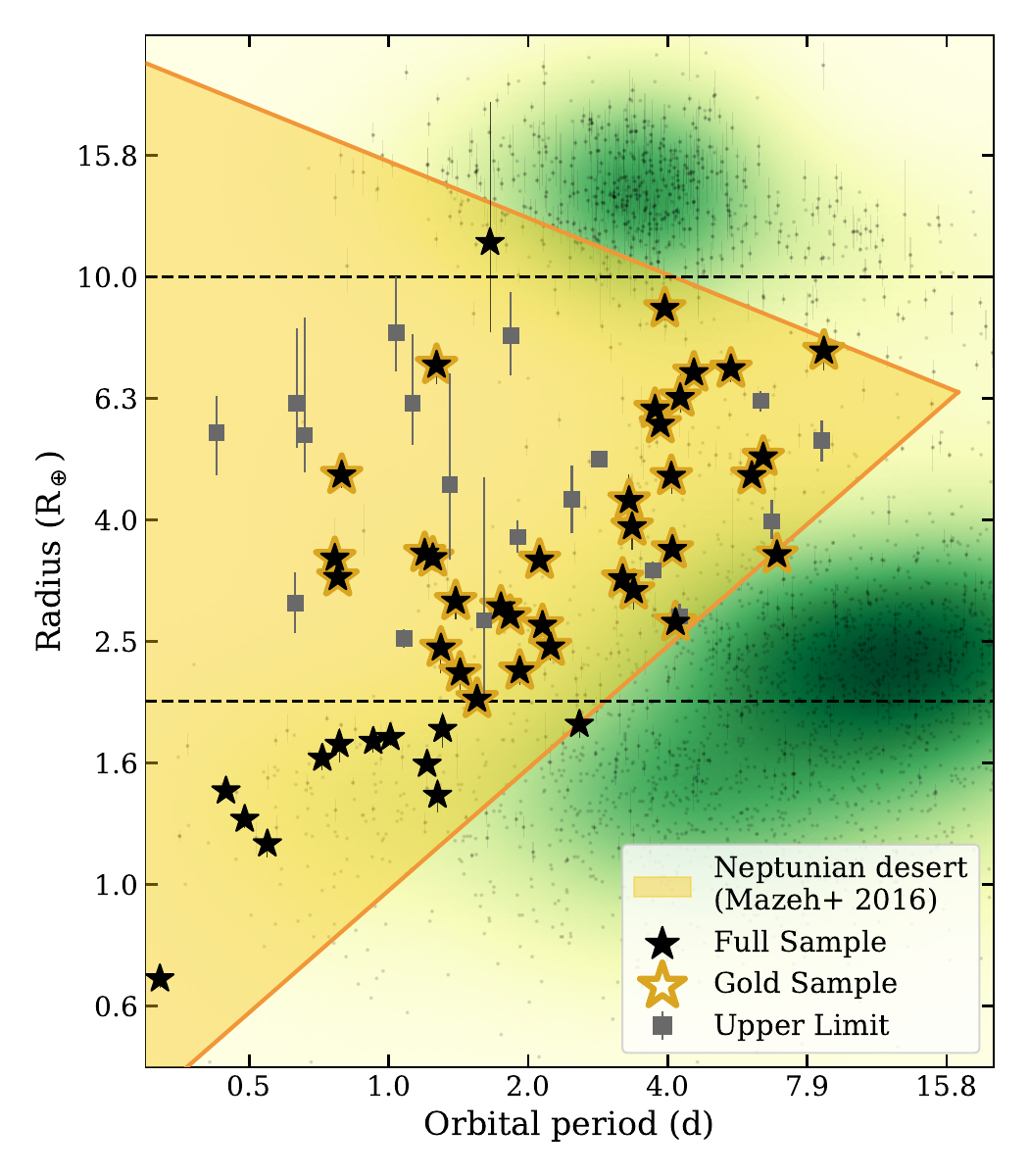}
    \caption{Our full sample of all 64 targets within the period-radius space (black stars). The Neptunian desert boundaries are from \citet{mazeh2016dearth} plotted as solid lines, with the enclosed desert area shaded in yellow. The dashed black lines represent the boundaries at 2~\Rearth and 10~\Rearth used to define the Neptune `gold' sample (black stars highlighted in gold). Known planets were sourced from the NASA exoplanet archive (\url{exoplanetarchive.ipac.caltech.edu}) accessed on 28 January 2025 and are plotted as grey dots in the background. The population density of known planets is shaded in green, where darker green denotes more planets discovered in that region of parameter space. }
    \label{fig:nep_desert_all}
\end{figure}

\subsection{Full Sample}
\label{sec:full_sample}
After following the processes listed in previous sections and ranking each target using the merit function, we make a cut at a merit function of 5 where the sample was complete (i.e. precise mass measurements were known), finalising a sample of 64 TOIs which form our sample. The full list of 64 close-in Neptune targets along with their properties and merit can be found in Table \ref{tab:planet_prop}. Stellar parameters and their source for each TOI can be found in Table \ref{tab:nomads_stellar_prop}. Our HARPS observations were used to `complete' this sample, making sure that each candidate has received RV observations of sufficient precision to characterise the planet, if it is real. We list the status of each target where `PUB' is published (31 planets), `IN' is observed but currently unpublished at the time of writing (15 planets, of which we are missing parameters for 6) and `UP' is upper limit (18 targets). Where a target is observed but unpublished, we use preliminary values where available (private comm.) and treat the system as a planet, but with unknown mass otherwise. In total, the sample contains of 46 planets and 18 candidates with upper limits. Their location within the \citet{mazeh2016dearth} desert can be seen in Figure \ref{fig:nep_desert_all}. 

\begin{table*}
    \centering
    \caption{The planetary properties for all targets in our Neptunian desert sample. From left to right, the columns represent the TESS Object of Interest (TOI) ID, the Publication status ('PUB' is published, 'IN' is in-prep at the time of writing and 'UP' is upper limit), the Merit function (as calculated in \S \ref{sec:merit}), planetary mass, planetary radius, orbital period, planetary bulk density (calculated as 95\% upper limits), eccentricity, and the literature references for all planetary properties within this table. This full table is available as supplementary material online.}
    \resizebox{0.99\textwidth}{!}{ 
    \begin{tabular}{lcccccccr}
    \hline 
    TOI	    & Status & Merit	&  M$_{\rm{p}}$ & R$_{\rm{p}}$	& $P_{\rm{orb}}$ & $\rho$	& $e$ & Reference \\
        &  & 	&  (M$_{\oplus}$) & (\Rearth)	&  (days) & (gcm$^{-3}$)&  & \\
    \hline
422.01	& UP	& 21.44	&  $\leq 2.98$	                  & $6.19\substack{+2.03\\-0.96}$ & 0.6333 &  $\leq 0.069$                   & 0	                              & This work \\
193.01	& PUB	& 19.72	&  $29.32\substack{+0.78\\-0.81}$ &	$4.72\substack{+0.23\\-0.23}$ & 0.7920 & $1.536 \pm 0.123$	           & $\leq 0.058$                           & \citet{jenkins2020toi193} \\
1967.01	& UP	& 14.5	&  $\leq 6.5$		                  & $5.54\substack{+0.83\\-0.83}$ &	0.4262 & $\leq 0.210$				       & 0                                & Jenkins et al. (priv. comm.) \\
855.01	& UP	& 12.22	&  $6.38 \pm 2.31$ 	              & $8.0\substack{+1.46\\-1.12}$  & 1.8301 & $0.069\substack{+0.012\\ -0.0002}$ & 0	                              & This work \\
728.01	& UP	& 12.05	&  $\leq 6.49$	                  & $2.72\substack{+1.96\\-0.68}$ &	1.6024 & $\leq 1.772$			           & 0	                              & This work \\
355.01	& UP	& 11.49	&  $\leq 7.51$	                  & $8.1\substack{+1.92\\-1.11}$  & 1.0373 & $\leq 0.078$	                   &	0	                          & This work \\
576.01	& PUB	& 11.40	&  $32.1 \pm 1.6$                 & $7.06 \pm 0.34$               &	5.4435 & $0.5453 \pm 0.09$			   & $\leq0.07$                         & \citet{hellier2019toi576} \\
465.01	& PUB	& 10.82	&  $40.7\substack{+3.2\\-2.9}$    & $5.72 \pm 0.22$	              & 3.8362 & $1.33 \pm 0.1$	               & $\leq 0.007$                           & \citet{demangeon2018toi465} \\
1975.01	& UP	& 10.75	&  $\leq 4.07$                      & $5.01 \pm 0.1$	              & 2.8319 & $\leq 0.178$                    & 0	                              & This work \\
118.01	& PUB	& 10.50	&  $16.6 \pm 1.3$                 & $4.71 \pm 0.17$               & 6.0360 & $0.87\substack{+0.12\\-0.11}$ & $\leq 0.07$                            & \citet{esposito2019toi118} \\
880.01	& IN	& 10.29	&  $22.806 \pm 0.789$             &	$5.05 \pm 0.22$	              &	6.3870 & $0.971\substack{+0.086\\-0.098}$ & 0	                              & Nielsen et al. (priv. comm.) \\
271.01	& UP	& 10.20	&  $\leq 2.41$	                  & $4.31\substack{+0.59\\-0.53}$ &	2.4759 & $\leq 0.17$		                   & 0	                              & \citet{armstrong2024ncores} \\
851.01	& UP	& 9.94	&  $\leq 6.37$	                  & $2.9\substack{+0.36\\-0.31}$  &	0.6294 & $\leq 1.435$                              & 0                                & This work \\
141.01	& PUB	& 9.87	&  $8.83\substack{+0.66\\-0.65}$  &	$1.745 \pm 0.52$              & 1.0080 & $9.15\substack{+1.1\\1.0}$    & $\leq 0.24$                            & \citet{espinoza2020toi141} \\
192.01	& PUB	& 9.70	&  $77.547 \pm	6.356$            & $8.878\substack{+0.63\\-0.39}$& 3.9227 & $0.65\pm 0.10$ 	           & 0.03                             & \citet{hellier2010toi192} \\
510.01	& UP	& 9.66	&  $\leq 2.22$	                  & $4.55\substack{+2.40\\-1.13}$ & 1.3523 & $\leq 0.13$                         & 0	                              & \citet{armstrong2024ncores} \\
2498.01	& PUB	& 9.6	&  $34.62\substack{+4.10\\-4.09}$ &	$6.06\substack{+0.29\\-0.27}$ &	3.7382 & $0.86\substack{+0.25\\-0.2}$  & $0.089 \pm 0.075$                 &	\citet{frame2023toi2498} \\
829.01	& IN	& 9.49	&  $21.666 \pm 1.56$              &	$4.28 \pm 0.44$               & 3.2870 & $1.513\substack{+0.31\\-0.47}$ & 0	                              & Nielsen et al. (priv. comm.) \\
2358.01	& UP	& 9.42	&  $\leq 20.1$                      &	$6.2\substack{+1.86\\-0.90}$ &	1.1258 & $\leq 0.463$                              & 0                                & This work \\
641.01	& UP	& 9.39	&  $\leq 1.58$                      &	$3.73\substack{+0.25\\-0.22}$ & 1.8930 & $\leq 0.17$                         & 0                                & \citet{armstrong2024ncores} \\
4524.01	& PUB	& 9.23	&  $7.42 \pm 1.09$                & $1.72 \pm 0.07$	              & 0.9261 & $8.06 \pm 1.53$               & 0                                & \citet{murgas2022toi4524} \\
431.02	& PUB	& 9.12	&  $3.07 \pm 0.35$                & $1.28 \pm 0.04$               & 0.4900 & $8.0 \pm 1.0$                 & $\leq 0.28$	                          & \citet{osborn2021toi431} \\
564.01	& PUB	& 8.78	&  $465.0\substack{+31.8\\-30.5}$ & $11.4\substack{+8.0\\-3.3}$   & 1.6511 & $1.7\substack{+3.1\\-1.4}$    & $0.072\substack{+0.083\\-0.050}$ & \citet{davis2020toi564} \\
4461.01	& UP	& 8.65	&  $3.93\substack{+1.68\\-1.63}$  &	$3.29 \pm 0.11$               & 3.7006 & $0.60\substack{+0.17\\-0.28}$ & 0	                              & This work \\
2673.01	& IN	& 8.50	&                                 & $2.25 \pm 0.12$               & 1.9126 &                               &                                  & ExoFOP \\
1943.01	& IN	& 8.35	&                                 & $2.87 \pm 0.19$ 	          & 1.7425 &                               &                                  & ExoFOP \\
5556.01	& PUB	& 8.31	&  $5.60 \pm 0.43$                & $1.61 \pm 0.07$	              & 0.7195 & $7.4 \pm 1.1$                 & 0		                          & \citet{frustagli2020toi5556} \\
181.01	& PUB	& 8.21	&  $46.17\substack{+2.71\\-7.83}$ & $6.95\substack{+0.09\\-0.10}$ &	4.5320 & $0.76\substack{+0.015\\-0.1}$ & $0.1543\substack{+0.06\\-0.03}$  & \citet{mistry2023toi181} \\
261.01	& IN	& 8.18	&                                 & $3.05 \pm 0.21$               & 3.3639 &                               &                                  & \citet{hord2024toi261} \\
132.01	& PUB	& 8.15	&  $22.40\substack{+1.90\\-1.92}$ & $3.42\substack{+0.13\\-0.14}$ &	2.1097 & $3.08\substack{+0.44\\-0.46}$ & $0.087\substack{+0.054\\-0.057}$ & \citet{diaz2020toi132} \\
426.01	& IN	& 7.71	&  $7.3 \pm	1.7$                  & $2.45 \pm 0.22$               & 1.2947 & $2.72\substack{+0.13\\-1.11}$  & 0                                & Castro-González et al. (in-prep) \\
824.01	& PUB	& 7.60	&  $18.467\substack{+1.84\\-1.88}$& $2.93\substack{+0.20\\-0.19}$ &	1.3929 & $4.03\substack{+0.98\\-0.78}$ & 0                                & \citet{burt2020toi824} \\
880.02	& IN	& 7.55	&  $3.059 \pm 0.573$              & $1.83 \pm 0.094$              & 2.5700 & $2.72\substack{+0.06\\-0.13}$ & 0                                & Nielsen et al. (priv. comm.) \\
969.01	& PUB	& 7.55	&  $9.1	\pm 1.0$	              & $2.765\substack{+0.09\\-0.10}$& 1.8237 & $2.34\substack{+0.39\\-0.34}$ & 0                                & \citet{lillo2023toi969} \\
849.01	& PUB	& 7.54	&  $39.09\substack{+2.66\\-2.55}$ &	$3.44\substack{+0.16\\-0.12}$ & 0.7655 & $5.5\substack{+0.7\\-0.8}$    & $\leq 0.08$                            & \citet{armstrong2020rtoi849} \\
4537.01	& UP	& 7.37	&  $\leq 11.2$                      &	$3.96\substack{+0.34\\-0.26}$ &	6.6612 & $\leq 0.99$                     & 0                                & This work \\
5632.01	& PUB	& 6.93	&  $18.8 \pm 2.2$                 &	$6.33\substack{+0.81\\-0.36}$ & 4.2345 & $0.40 \pm 0.1$                & $0.124 \pm 0.06$                 & \citet{hartman2011toi5632} \\
561.02	& PUB	& 6.91	&  $2.0 \pm 0.23$                 & $1.43 \pm 0.037$              &	0.4465 & $3.8 \pm 0.5$                 & 0                                & \citet{weiss2021toi561} \\
&&&&&&&&\citet{lacedelli2022toi561} \\
2427.01	& IN	& 6.77	&                                 & $1.8 \pm 0.12$                &	1.3060 &                               &                                  & \citet{giacalone2022toi2427} \\
745.01	& UP	& 6.73	&  $\leq2.51$                       &	$2.54 \pm 0.09$	              & 1.0791 & $\leq 0.84$                     & 0                                & \citet{armstrong2024ncores} \\
2365.01	& UP	& 6.68	&  $\leq 8.11$                      & $5.49\substack{+3.08\\-0.72}$ & 0.6597 & $\leq 0.27$	                   & 0                                & This work \\
5005.01	& UP	& 6.48	&  $32.7 \pm 5.9$   	          & $6.25 \pm 0.24$	              & 6.3085 & $0.74 \pm 0.16$               & 0                                & \citet{castro2024toi5005} \\
3071.01	& PUB	& 6.29	&  $68.2 \pm 3.5$                 &	$7.16\substack{+0.57\\-0.51}$  &	1.2669 & $1.02\substack{+0.26\\-0.22}$ & $\leq 0.09$                            & \citet{hacker2024toi3071} \\
1839.01	& IN	& 6.21	&  $6.3 \pm	0.9$	              & $2.23 \pm 0.16$               & 1.4237 & $3.12 \pm 0.22$	           & 0                                & Castro-González et al. (in-prep) \\
5559.01	& IN	& 6.18	&  $11.1\substack{+1.5\\-1.4}$    &	$2.67 \pm 0.05$               &	2.1408 & $3.2\substack{+0.5\\-0.4}$    & $\leq 0.16$                            & \citet{sozzetti2024toi5559} \\
731.01	& PUB	& 6.04	&  $0.633 \pm 0.05$	              &	$0.699 \pm 0.024$             &	0.3219 & $10.2 \pm 1.3$	               & $0.060\substack{+0.07\\-0.04}$   & \citet{lam2021toi731} \\
&&&&&&&&\citet{goffo2023toi731} \\
1117.01	& IN	& 6.01	&  $8.90\substack{+0.95\\-0.96}$  &	$2.46\substack{+0.13\\-0.12}$ &	2.2281 & $3.30 \pm 0.6$	               & 0                                & Lockley et al. (submitted) \\
1853.01	& PUB	& 6.00	&  $73.5\substack{+4.2\\-4.0}$    & $3.45\substack{+0.13\\-0.14}$ & 1.2436 & $9.8\substack{+0.82\\-0.76}$  & $\leq 0.03$                            & \citet{naponiello2023toi1853} \\
544.01	& PUB	& 5.96	&  $2.89 \pm 0.48$	              & $2.018 \pm	0.076$            & 1.5483 & $1.93\substack{+0.30\\-0.25}$ & $0.35\substack{+0.14\\-0.12}$    & \citet{osborne2024toi544} \\
499.01	& UP	& 5.81	&  $\leq 6.66$                      & $5.38\substack{+0.43\\-0.41}$ & 8.5334 & $\leq 0.24$                     & 0                                & This work \\
2196.01	& PUB	& 5.79	&  $26.0 \pm 1.3$                 & $3.51 \pm 0.15$               &	1.1947 & $3.31\substack{+0.51\\-0.43}$ & 0                                & \citet{persson2022toi2196} \\
179.01	& PUB	& 5.74	&  $25.5 \pm 4.6$ 	              &	$2.7 \pm 0.05$	              & 4.1374 & $7.1 \pm 1.4$                 & $0.2\substack{+0.1\\-0.2}$	      & \citet{vines2023toi179} \\
    \hline
    \end{tabular}}
    \vspace{-2mm}
     \begin{flushleft}
   {\bf Notes:}  The Exoplanet Follow-up Observing Program (ExoFOP) website serves as a repository for project and community-gathered data by allowing upload and display of data and derived astrophysical parameters. For targets with parameters from ExoFOP it can be accessed directly at \url{https://www.ExoFOP.ipac.caltech.edu/tess/}. A description of derived parameters from this paper can be found in Appendix \ref{sec:null_results}. 
    \end{flushleft}
    \label{tab:planet_prop}
\end{table*}

\begin{table*}
    \centering
    \contcaption{The planetary properties for all targets in our Neptune desert sample.}
    \resizebox{0.99\textwidth}{!}{ 
    \begin{tabular}{lcccccccr}
    \hline 
    TOI	    & Status & Merit	&  M$_{\rm{p}}$ & R$_{\rm{p}}$	& $P_{\rm{orb}}$ & $\rho$	& $e$ & Reference \\
        &  & 	&  (M$_{\oplus}$) & (\Rearth)	&  (days) & (gcm$^{-3}$)&  & \\
    \hline
332.01	& PUB	& 5.7	&  $57.2 \pm 1.6$                 & $3.2\substack{+0.16\\-0.11}$  &	0.7770 & $9.6\substack{+1.1\\-1.3}$    & $\leq 0.15$                            & \citet{osborn2023toi332} \\
500.01	& PUB	& 5.68	&  $1.42 \pm 0.18$                &	$1.17 \pm 0.06$	              & 0.5481 & $4.89\substack{+1.03\\-1.88}$ & $0.063\substack{+0.068\\-0.044}$ & \citet{serrano2022toi500} \\
2350.01	& IN	& 5.62	&                                 & $7.55 \pm 0.55$               &	8.6251 &                               &                                  & ExoFOP \\
238.01	& PUB	& 5.57	&  $3.40 \pm 0.46$	              & $1.40 \pm 0.09$	              & 1.2730 & $6.8 \pm 1.6$                 & $\leq 0.18$                            & \citet{mascareno2024toi238} \\
4517.01	& PUB	& 5.49	&  $5.14 \pm 0.47$	              & $1.58 \pm 0.03$               & 1.2089 & $7.19\substack{+0.81\\-0.76}$ & $\leq 0.063$                           & \citet{kosiarek2020toi4517} \\
&&&&&&&&\citet{bonomo2023toi4517} \\
1130.02	& PUB	& 5.48  &  $19.28 \pm 0.97$               &	$3.56 \pm 0.13$               &	4.0744 & $2.34 \pm	0.26$              & $0.0541 \pm 0.0015$              & \citet{korth2023toi1130} \\
908.01	& PUB	& 5.3	&  $16.14\substack{+4.1\\-4.04}$  &	$3.19 \pm 0.16$	              &	3.1837 & $2.74\substack{+0.24\\-0.35}$ & $0.132 \pm	0.091$                & \citet{hawthorn2023toi908} \\
5094.01	& PUB	& 5.29	&  $12.14\substack{+0.68\\-0.66}$ & $3.88 \pm 0.32$               &	3.3366 & $0.93\substack{+0.56\\-0.31}$ & $0.123 \pm	0.04$                 & \citet{biddle2014toi5094}  \\
&&&&&&&&\citet{stefansson2022toi5094} \\
2411.01	& IN	& 5.23	&  	                              & $1.74 \pm 1.6$                & 0.7826 &                               &                                  & ExoFOP \\
442.01	& PUB	& 5.21	&  $30.8 \pm 1.5$                 & $4.7 \pm 0.3$                 & 4.0520 & $1.7 \pm 0.3$	               & $0.04 \pm 0.02$	              & \citet{dreizler2020toi442} \\
682.01	& IN	& 5.21	&  $12.29\substack{+1.0\\-0.99}$  &	$3.49\substack{+0.12\\-0.11}$ & 6.8396 &  $1.58\substack{+0.04\\-0.02}$                             & $0.411\substack{+0.033\\-0.031}$ & \citet{quinn2021toi682} \\
2227.01	& UP	& 5.03	&  $\leq 10.9$                      & $2.76\substack{+0.14\\-0.13}$ & 4.2217 & $\leq 2.84$                     & 0                                & This work \\
    \hline
    \end{tabular}}
    \vspace{-2mm}
     \begin{flushleft}
   {\bf Notes:}  The Exoplanet Follow-up Observing Program (ExoFOP) website serves as a repository for project and community-gathered data by allowing upload and display of data and derived astrophysical parameters. For targets with parameters from ExoFOP it can be accessed directly at \url{https://www.ExoFOP.ipac.caltech.edu/tess/}. A description of derived parameters from this paper can be found in Appendix \ref{sec:null_results}. 
    \end{flushleft}
\end{table*}

\subsection{The `gold' sample}
\label{sec:gold_sample}
Our full sample detailed in \S \ref{sec:full_sample} utilises the \citet{mazeh2016dearth} desert boundaries in period-radius space which cover a wide range in planetary radius, beyond the Neptunian domain. Therefore, to focus on Neptune-sized planets, we define a subset of our full sample which are present within the \citet{mazeh2016dearth} boundaries and have more constrained radii. We place boundaries in radius at 2~R$_{\oplus}$ and 10~R$_{\oplus}$ on the period-radius diagram, selecting all targets which fall inside. Figure \ref{fig:nep_desert_all} shows these boundaries and highlights the 33 planets which make up our `gold' sample. From our 33 planets, four (TOI-2673.01, TOI-1943.01, TOI-261.01 and TOI-2350.01) do not have mass measurements in the literature (status `IN' in Table \ref{tab:planet_prop}, with the analyses still ongoing by other groups, private comm.). As such, they are treated as unconstrained in any analysis where the planet mass is required, but remain in the sample for studying host star metallicity and multiplicity. 

\subsection{Gaia DR3 Cross-match}
\begin{figure}
    \centering
    \includegraphics[width=0.47\textwidth]{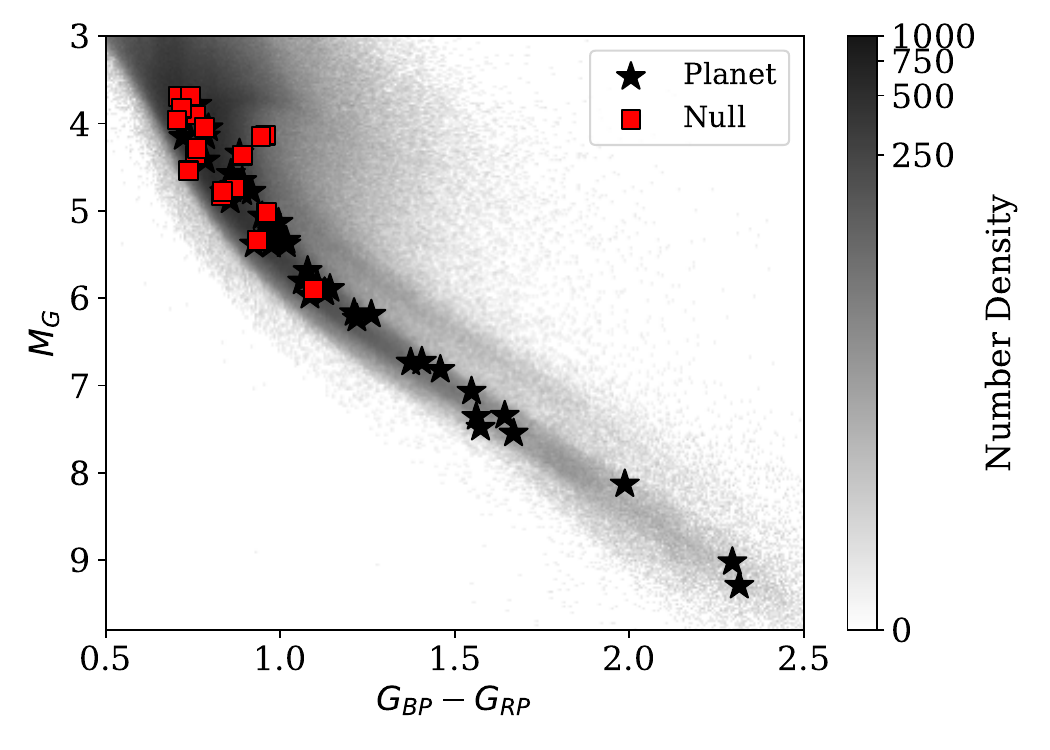}
    \caption{The colour-magnitude diagram of all 64 targets from our full sample in the Neptunian Desert sample, where black star markers indicate a planet, and red squares a null detection (i.e. no planet/false positive). The grey background sample represents the TESS-SPOC FFI Targets in TESS Sectors 1 – 55 taken from \citet{doyle2024tess}, where the colour scale represents the log of the density of stars. All colours along with the parallax, which are used to determine the absolute Gaia magnitude, were taken from the Gaia DR3 catalogue.}
    \label{fig:gaia_hrd}
\end{figure}

We crossed matched all 64 of targets from our full sample with Gaia DR3 to provide further stellar properties. As a result, we were able to plot all our targets on a colour-magnitude diagram which is shown in Figure~\ref{fig:gaia_hrd}. From this, it can be seen that the confirmed planets within our sample are spread across the main sequence. In contrast, all of the null results (TOIs that have been followed up and do not show a RV signal, and hence are more likely to be a false positive), are clustered around the top of the main sequence. This area of the main sequence represents stars which have late-A and F spectral types, with an increased fraction of these being members of binary systems \citep{tian2018binary} or even triple systems \citep{offner2023origin}. Overall, this suggests TOIs or other potential exoplanet candidates in this region of the colour-magnitude diagram have a higher likelihood of being a false positive which may be important to consider when performing ground-based follow-up. 

\section{Stellar and Planetary Companions}
\label{sec:multiplicity}

Theoretical models of planet formation and evolution primarily focus on single stars \citep[see][and references therein]{mordasini2024planet}. However, over half of the stars in our Galaxy are members of binary or higher order multiple systems \citep{raghavan2010survey}; therefore, considering the effects stellar companions have on exoplanet formation is essential to form a complete overview of the origins of exoplanetary systems. The gravity of a companion star can strongly influence the formation and dynamic evolution of emerging planets. Overall, the architecture of a system depends on how the efficiency of accretion processes are affected by the presence of a close massive body. Additionally, a second star will have effects on the evolution of the system through tidal interactions, migration, planet-planet scattering or Kozai mechanisms \citep[see][and references therein]{marzari2019planets}. 

There is a dearth of exoplanets seen in systems with close binary companions \citep[separations of $\lesssim$~50 AU,][]{kraus2016impact}. Theoretical studies have shown close binary companions can prevent planet formation by stirring up protoplanetary discs \citep[e.g.][]{mayer2005gravitational}, tidally truncating discs \citep[e.g.][]{pichardo2005circumstellar, kraus2011role} or leading to the ejection of planets \citep[e.g.][]{kaib2013planetary, zuckerman2014occurrence}. On the other hand, wider binary companions (separations of $\gtrsim$ 200~AU) are less detrimental to the survival of exoplanets and may even help planet migration processes \citep{moe2021impact}. For example, an outer companion can create regions of dense gas and particles in disk spiral arms, creating favourable conditions for planetesimals and pebble accretion \citep{youdin2005streaming, johansen2007rapid, lambrechts2014forming}. Additionally, considering the evolution of these planets, an outer companion could cause secular interactions leading to inward migration \citep{naoz2014mergers}. For giant planets and brown dwarfs above 7~M$_{\rm{J}}$ and within 1~AU, \citet{fontanive2019high} found 80\% have a host star with an outer binary companion between 20 - 10,000~AU. Overall, it is clear stellar companions play a role in the existence and properties of giant planets; however, the precise nature and extent of their role is not fully understood. 

In addition to stellar companions, we have to also consider the effects that planetary companions within a system have on each other. Simultaneous planets forming within a disk can influence planet formation through interactions (either planet-planet or planet-disk) which ultimately decide the fate of the final planetary architecture \citep[e.g.][and references therein]{voelkel2022exploring}. Within our own solar system there is a clear divide between terrestrial planets and gas giants, suggesting they formed and evolved separately from each other or the presence of Jupiter sculpted the inner solar system. Exoplanet systems similar to our solar system appear rare \citep[e.g. see][]{cabrera2013planetary}, with this believed to be a result of larger planets tending to migrate inwards \citep{kley2012planet} or observational biases in detecting long period transiting gas giants \citep{weiss2023kepler}. Multi-planet systems hosting large gas giants on short orbital periods tend to have no close-in smaller terrestrial planets due to this migration \citep{dawson2018origins}.  

\begin{figure}
    \centering
    \includegraphics[width=0.47\textwidth]{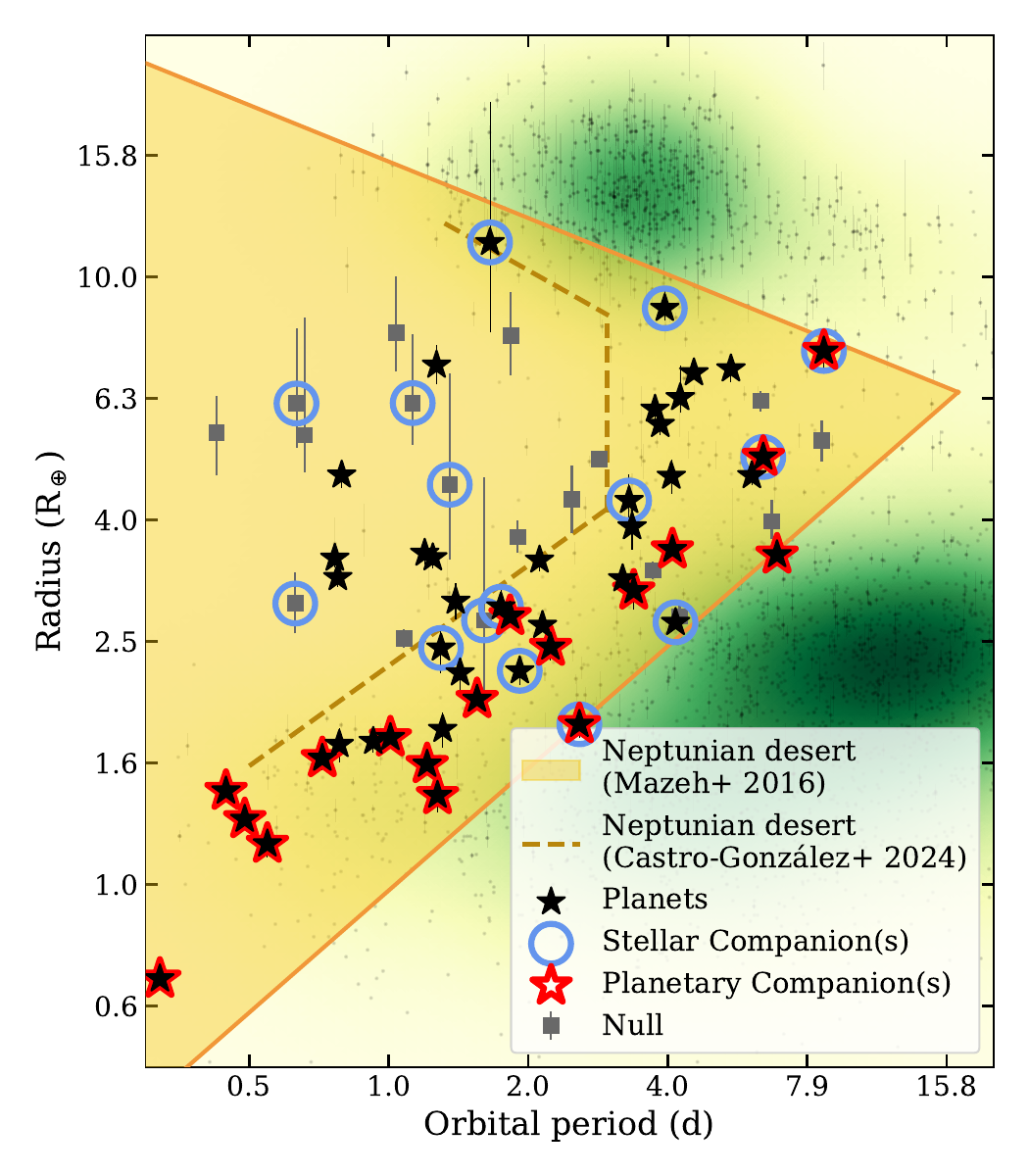}
    \caption{Our full sample of all 64 targets within the period-radius space. Targets identified as hosting a stellar companion (i.e. Stellar Companion(s)) are plotted with a light blue circle. Those which are members of a multiple planet system (i.e. Planetary Companion(s)) are plotted with a red highlight. Some targets meet both criteria. The Neptunian desert boundaries from \citet{mazeh2016dearth} are plotted as solid lines, with the enclosed Neptunian desert area shaded in yellow. The updated Neptunian desert boundaries from \citet{castro2024mapping} are plotted as a golden dashed line. All known planets were sourced from the NASA exoplanet archive (\url{https://www.exoplanetarchive.ipac.caltech.edu}) accessed on 28 January 2025 and are plotted as grey dots in the background. The population density of known planets is shaded in green, where darker green denotes more planets discovered in that region of parameter space. }
    \label{fig:companions}
\end{figure}

\begin{figure}
    \centering
    \includegraphics[width=0.47\textwidth]{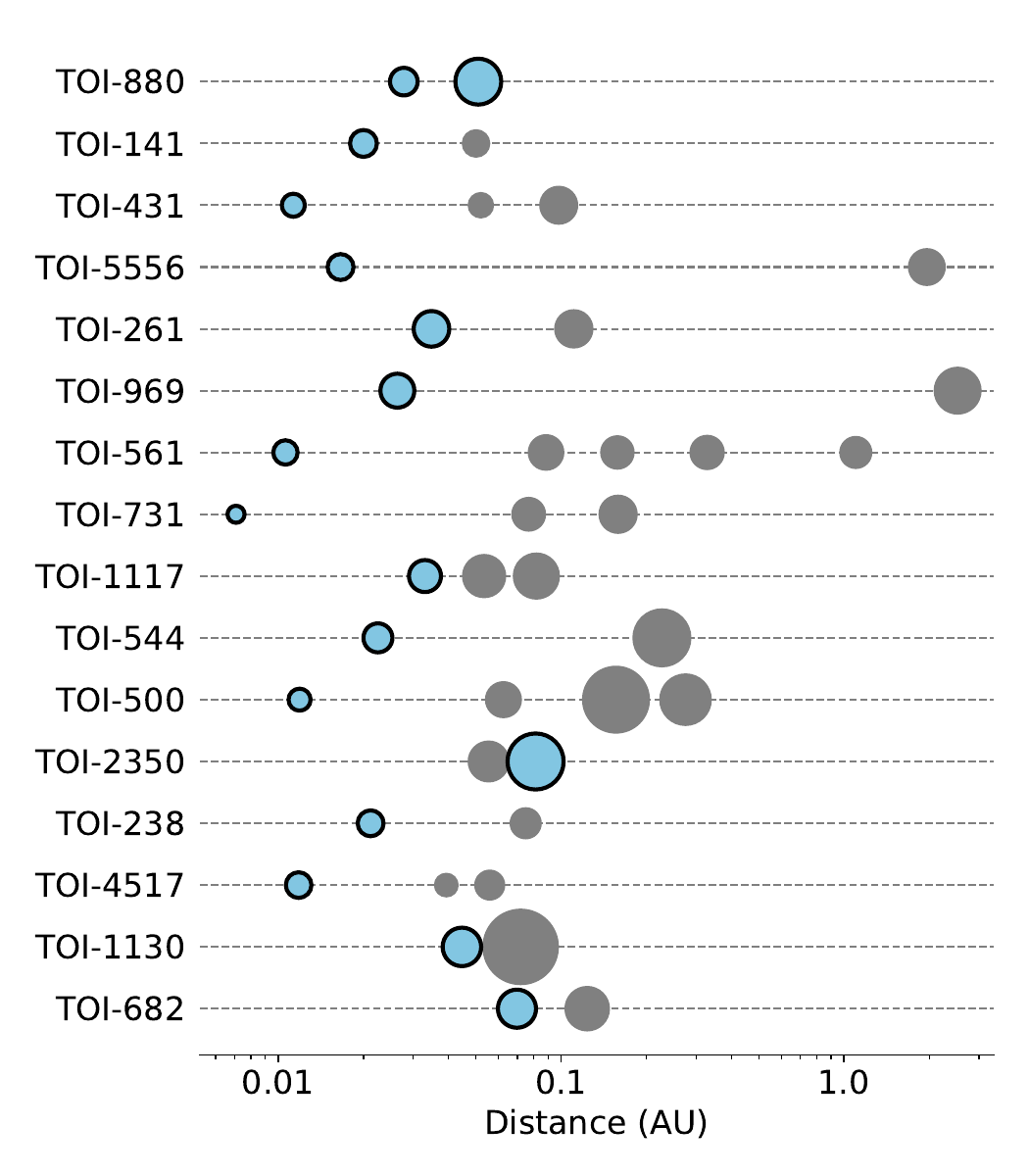}
    \caption{A schematic showing the configurations of each target identified as being a member of a multi-planet system. Planets coloured blue with a black outline are the targets in our sample and all other planets in the system are in grey. For planets in the multi-planet system where only an orbital period and projected planetary mass were given in the literature, we estimated the planetary radius using the mass-radius relation of \citet{muller2024mass} in equation 5 (assuming $M_{\rm{p}}\sin i_{\star} = M_{\rm{p}}$). We calculated the semi-major axis using $a = (P^2 G M_{\star}/4\pi^2)^{1/3}$. These properties were only used for the creation of this plot as a visual aid. }
    \label{fig:multi-planets}
\end{figure}

\begin{table}
    \centering
    \caption{The stellar host and companion properties of all targets from our sample which have been identified with a stellar companion. The columns from left to right represent the TESS Object of Interest (TOI) ID, Renormalised Unit Weight Error (RUWE) from the Gaia DR3 catalogue, the distance to the host star estimated as the inverse of the Gaia parallax, and the separation of the binary and the companion mass, both taken from the The Multiplicity of {\tess} Objects of Interest catalogue \citep{mugrauer2020gaia, mugrauer2021gaia, mugrauer2022gaia, mugrauer2023gaia}.}
    \begin{tabular}{lcccc}
    \hline
         TOI & RUWE & distance & binary separation & companion mass \\
             &      & (pc)     & (AU)              & $M_{\odot}$    \\
    \hline
422  &  0.979 &	124.1	&   3736	& 1.41 \\
728  &  1.104 &	172.2	&   375	    & 0.7  \\
880  &  0.996 &	60.56	&   503	    & 0.25 \\
851  &  2.151 &	169.6	&   287	    & 0.2  \\
192  &  0.970 &	87.45	&   10994	& 0.38 \\
510  &  2.432 &	92.84	&   512	    & 0.74 \\
829  &  1.060 &	138.9	&   781	    & 0.4  \\
2358 &  1.030 &	383.3	&   2802	& 0.73 \\
564  &  1.129 &	198.8	&   100	    & 0.3  \\
2673 &  1.335 &	99.19	&   106	    & 0.66 \\
1943 &  0.799 &	130.49	&   473	    & 0.5  \\
426  &  1.172 &	112.8	&   1012	& 0.85 \\
179  &  0.993 &	38.63	&   3376	& 0.72 \\
2350 &  1.365 &	187.1	&   5874	& 0.2  \\
    \hline     
    \end{tabular}
    \label{tab:binary}
\end{table}

\begin{figure*}
    \centering
    \includegraphics[width=0.97\textwidth]{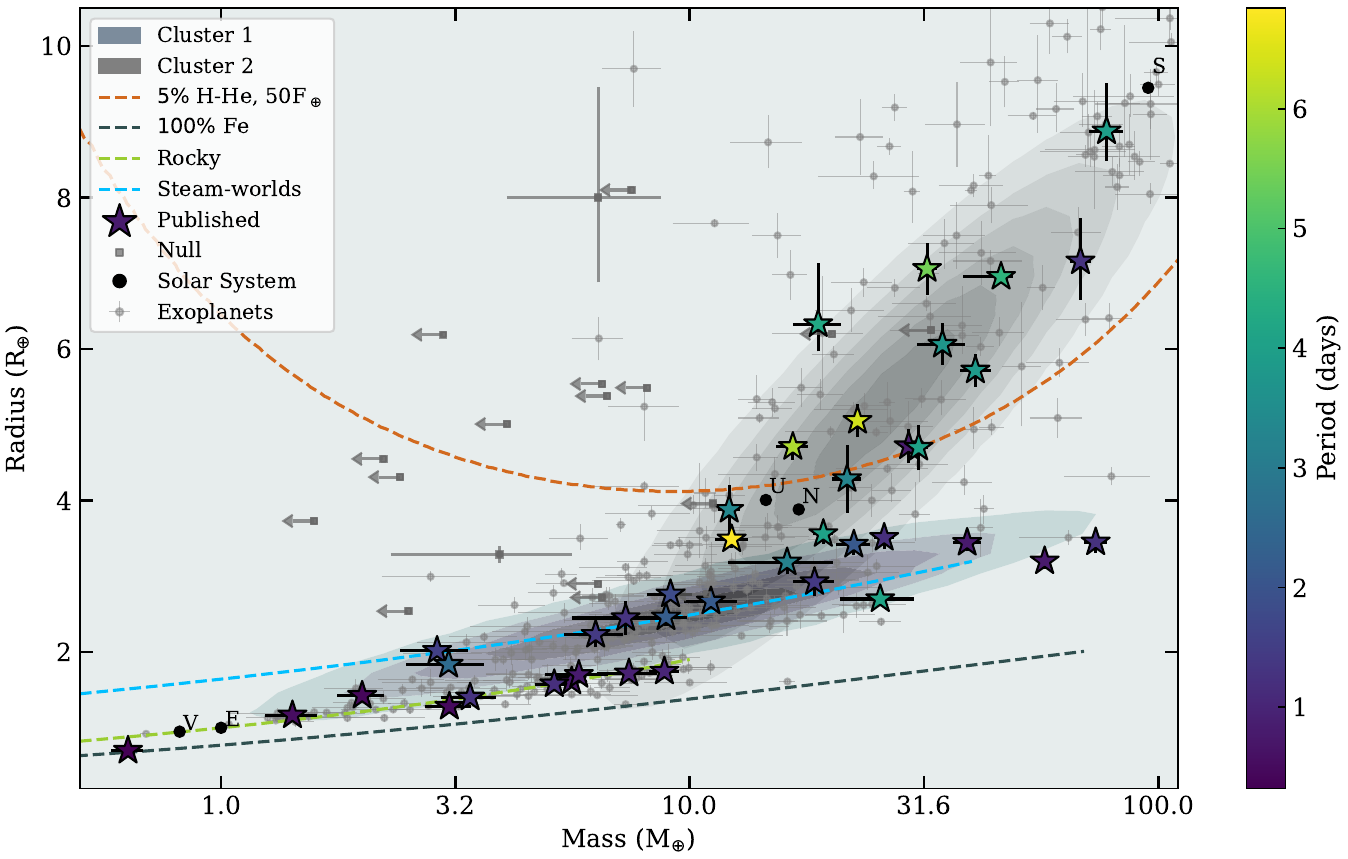}
    \caption{Mass-radius diagram for all 64 targets in our full sample, including confirmed planets (star markers) and null detections (square markers). The colour of each of the planet markers indicates their orbital period. The dashed lines are the compositions for pure iron (dark grey), plus an additional line representing a 5 per cent H-He envelope with $F/F_{\oplus}$ = 50 (red) \citep[taken from][]{otegi2020revisited}. Other composition lines include Earth-like (green), and Steam-world (blue) \citep[taken from][]{venturini2024fading}. The Solar System planets Venus (V), Earth (E), Uranus (U), Neptune (N) and Saturn (S) are shown for context as black circles. The background light grey exoplanet sample  is taken from the NASA Exoplanet archive (\url{https://www.exoplanetarchive.ipac.caltech.edu}, as of 28 January 2025) with  mass determinations better than 4$\sigma$. TOI-564~b is not shown in this Figure for visual representation of the whole sample, as it has a mass of $M_{\rm{p}}$ = 465~M$_{\oplus}$ and radius $R_{\rm{p}}$ = 11.4~R$_{\oplus}$ of putting it outwith the plotted region.}
    \label{fig:mass_radius_all}
\end{figure*}

We investigate if any of our Neptune desert targets from our full sample have either stellar and/or planetary companions. The stellar or planetary multiplicity is flagged according to the corresponding reference provided for each system in Table \ref{tab:planet_prop}. We then also used The Multiplicity of {\tess} Objects of Interest catalogue \citep{mugrauer2020gaia, mugrauer2021gaia, mugrauer2022gaia, mugrauer2023gaia} to confirm or determine the stellar multiplicity for all hosts in our sample. The results of our search are plotted in Figure \ref{fig:companions}, where we show our sample in radius-period space with light blue circles indicating the presence of a stellar host companion and dark blue circles indicating the presence of planetary companions. Only three targets have both stellar and planetary companions and these are TOI-880.01, TOI-880.02 and TOI-2350.01. In this case these targets have been colour coded according to both binarity of the stellar host and the multiple planetary system, where the latter appears as highlighted in red, see Figure \ref{fig:companions}. 

Overall, a total of 15 (23\%) planets in our full sample have stellar companions (Table \ref{tab:binary}), and these appear to be scattered across the radius-period space. All of these companions have separations which are greater than 100~AU, with six having separations of thousands of AU. Therefore, from ourfull sample of 64 targets, there appears to be evidence of no strong correlation between whether the planet's host star had a companion and its location in the period-radius space. However, targets with stellar companion separations in the hundreds of AU could influence inward migration of orbiting planets. Within our sample we have a handful of targets with stellar companions at separations of a few hundred AU, where the planetary orbital period is $<$~2~days, tentatively supporting the inward migration of orbiting planets. With regards to planetary companions, a total of 17 (26\%) of targets in our full sample are members of a multi-planet system, where all of these lie towards the lower boundary of the \citet{mazeh2016dearth} desert. Interestingly, the revised desert boundaries by \citet{castro2024mapping} show a more restrictive lower boundary, which leaves out all the planets with planetary companions (see Fig.~\ref{fig:companions}). Therefore, according to these new boundaries, no Neptune desert planet in our sample was found in multiplanet systems. 

We plot the configuration of each multi-planetary system in the schematic shown in Figure \ref{fig:multi-planets} to give some context. Only one of the targets in our sample has an interior, non-desert planet, suggesting that for the majority of the sample any smaller interior planets have been removed or ejected as a result of high eccentricity inward migration of the sample planet. Additionally, inner planets to gas giants may have gone undetected due to their small mass and resulting low amplitude RV signal. There are two exceptions to this picture; firstly TOI-880, which hosts two planets in our sample. Their orbits are in close proximity, and they could have potentially migrated as a pair. The second exception is TOI-2350, which also has two gas giant planets which could have migrated as a pair; however, masses for these have not yet been determined so this could change the landscape of the system. Furthermore, both TOI-880 and TOI-2350 have stellar companions at wide binary separations, which could have influenced the inward migration of these planets as pairs \citep{naoz2014mergers}. On the other hand, a handful of systems have larger planets (in radius) on exterior orbits which could have influences on the inward migration of the interior target, similar to a wide binary companion \citep{naoz2014mergers}. 

\begin{figure}
    \centering
    \includegraphics[width=0.47\textwidth]{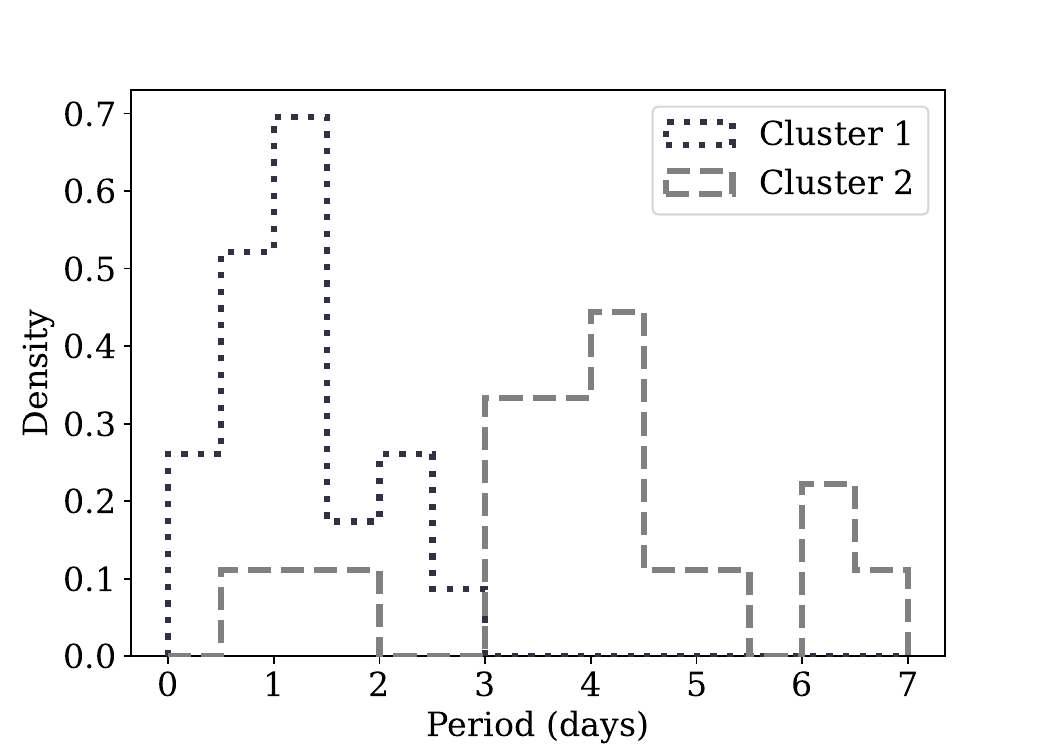}
    \caption{The orbital period distributions for the two clusters (1 as blue-grey dotted and 2 as grey dashed lines) identified in the mass-radius space of our Neptune desert sample. This shows a tentative split in orbital period at $\sim$3~days. }
    \label{fig:period_distribution}
\end{figure}


\section{Density}
\label{sec:density}
We plot our full sample of 64 targets in mass-radius space, shown in Figure \ref{fig:mass_radius_all}, with points colour coded according to orbital period, $P_{\rm{orb}}$. From this, we can see a tentative trend where planets with shorter periods cluster around the Earth-like and water lines, including notably the larger 3 -- 4$R_{\oplus}$ short period planets. Longer period planets typically sit above the water and H-He lines. This suggests photo-evaporation is at play for these short period planets where their planetary atmosphere is evaporated \citep[e.g.][]{owen2018photoevaporation}. To quantify this we run a Gaussian Mixture Model \citep[GMM;][]{scikit-learn} which is a function composed of several Gaussians and can be used to separate datapoints into clusters. In the case of our sample, we fit the mass, radius and orbital period of all 64 of our Neptune desert targets in our full sample with a 1, 2 and 3-component GMM, comparing the BIC of each model to determine which fits our data best. Overall, a difference in BIC of $\sim$6 between models signifies strong evidence of the lower BIC model being the better fit to the data \citep{raftery1995bayesian,lorah2019value}. The $\Delta$BIC for each model where we set the 2-component model as zero is as follows; (i) 1-component, $\Delta$BIC = 20.53, (ii) 2-component, $\Delta$BIC = 0.0 and (iii) 3-component, $\Delta$BIC = 52.99. Therefore, the 2-component GMM fits our data best and separates our sample into two clusters which independently match the short and long-period split observed. The resulting clusters are shown as shaded regions in Figure \ref{fig:mass_radius_all}. From this, we can then plot the distribution of orbital period in each of our clusters, shown in Figure \ref{fig:period_distribution}, which shows a tentative split in orbital period at $\sim$3~days. Overall, this reinforces the suggestion that planets in this region show a split in composition by orbital period and hence, there are two formation and evolution mechanisms at play.


\section{Envelope Mass Fractions}
\label{sec:emf}
The knowledge of both planetary mass and radius enables the determination of the planetary bulk density, $\rho$, which allows for possible compositions and internal structures of planets to be inferred. The planetary bulk density is the result of not only composition, but also irradiation and mass. As H-He has a low density, a small amount (e.g. 1\% of a planet's total composition) has a high impact on the radius of a Neptunian planet, becoming a dominant contribution to its size. However, it can be more informative to consider the interior structures of each planet. 

Interior structure modelling is required to obtain accurate Envelope Mass Fractions (EMF). Therefore, we perform an analysis with interior-atmosphere models to calculate the EMF instead of using the bulk density as a proxy of envelope mass. The following sections detail our interior structure analysis (\S \ref{sec:gastli}) and results (\S \ref{sec:emf_results}). 

For the interior structure models, we utilise the Neptune `gold' sample of 33 targets detailed in \S \ref{sec:gold_sample}. We require the planetary mass and planetary radius to estimate the composition, which are listed in Table \ref{tab:planet_prop}.Therefore, planets TOI-2673.01, TOI-1943.01, TOI-261.01 and TOI-2350.01 are not included in this analysis, resulting in 29 of our 'gold' sample targets for interior structure modelling. Additionally, the models also require the planetary equilibrium temperature, $T_{\rm{eq}}$ which are listed in Table \ref{tab:EMF}. The equilibrium temperatures for each target are calculated using the stellar effective temperature, $T_{\rm{eff}}$ and semi-major axis, $a$, using the following formula from \citet{acuna2024gastli}:

\begin{equation}
    T_{\rm{eq}}^4 = \frac{T_{\rm{eff}}}{f_{av}} \left(\frac{R_{\star}}{a}\right)^2,
\end{equation}

where the factor $f_{av} = 4$ as the equilibrium temperature is assumed to be a global average \citep[i.e. we assume the irradiation received from the star is fully distributed across all latitudes and longitudes in the planet, see][]{molliere2015model} and, $R_{\star}$ is the stellar radius. 

\subsection{Interior structure modelling}
\label{sec:gastli}
We use the GAS gianT modeL for Interiors \citep[GASTLI:][]{acuna21_interiors,acuna2024gastli} to calculate the EMF for 21 of  our `gold' sample and \citep{lopez2014understanding} for the remaining 8 targets. This is due to these planets being too dense and not massive enough to be gas giants and, therefore, are in the sub-Neptune regime not covered by GASTLI. The GASTLI model consists of two layers: an innermost core composed of a 1:1 mixture of rock and water, and an outer envelope containing H/He and water (collectively referred to as 'metals'). GASTLI is applicable across a broad range of planetary masses, provided that the atmospheric grid used for coupling covers relevant surface gravities. The default atmospheric grid is suitable for masses in the range 17~M$_{\oplus}$ < $M_{\rm{p}}$ < 6~M$_{\rm{J}}$. This allows us to draw consistent conclusions from our EMF estimates, as using a single interior structure model for all targets creates a homogeneous sample of calculated EMFs. For the 8 targets not covered by GASTLI the resulting EMFs are all near zero (Table \ref{tab:EMF}) as expected for denser planets, and so any model dependence would not affect the conclusions.

To estimate the EMF, we generated a grid of interior models that yield radius and age over varying mass, equilibrium temperature, core mass fraction (CMF), and internal temperature ($T_{int}$). For simplicity, the envelope metallicity was set to solar in all models, corresponding to an envelope metal mass fraction of 1.3\%. We employed the emcee package \citep{emcee_ref} to perform Markov chain Monte Carlo (MCMC) retrievals for each planet. To calculate the likelihood, we adopted equations 6 and 14 from \cite{dorn15_interiors} and \cite{acuna21_interiors}, respectively. The mass and radius are used as observables, but the age is not included as an observable as the age uncertainties of our `gold' sample are poorly constrained. Our priors are uniform for CMF = $\mathcal{U}(0,0.99)$, and $T_{int} = \mathcal{U}(50,300)$ K, while the mass follows a Gaussian prior consistent with the observed mean and uncertainties. The final EMF is derived from the CMF posterior distribution function as EMF = 1 - CMF.

Some of the planets in our `gold' sample have equilibrium temperatures exceeding the maximum $T_{eq}$ = 1000 K covered by GASTLI's default atmospheric grid. For these cases, we set $T_{eq}$ = 1000 K in our retrievals and apply a correction to the CMF to account for the irradiation effect. This correction is derived from the mass-radius relations at varying irradiation levels, and total and core masses in \cite{Fortney07}. We estimate that the inferred CMF difference between  models at 900 K and 1900 K for the same radius and age is $\Delta$CMF $(\Delta T_{eq} = 1000 \ K) = $ 0.1. For each individual planet, this correction is scaled linearly with equilibrium temperature as per Eq. \ref{eq:cmf_scaling}, then applied according to Eq. \ref{eq:cmf_correction}.

\begin{equation} 
\label{eq:cmf_scaling}
    \Delta CMF (T_{eq}) = \dfrac{(T_{eq} - 1000) \times 0.1}{1000}
\end{equation}

\begin{equation} 
\label{eq:cmf_correction}
    CMF_{corrected} = CMF_{1000 \ K} +  \Delta CMF (T_{eq}).
\end{equation}

\begin{table}
    \centering
    \caption{Parameters of the 29 `gold' Neptunian desert sample which are required for the determination of Envelope Mass Fractions (EMF). The columns from left to right represent the TESS Object of Interest (TOI) ID, equilibrium temperature, semi-major axis, stellar age and EMF as calculated using GASTLI as described in \S \ref{sec:gastli} or models from \citet{lopez2014understanding} (indicated by a $^{*}$). For the EMF values, if the retrieval was able to reproduce the PDF of the radius and mass that are observed, then the values are given with errors. If the retrieved radius is larger than the observed radius, then the PDF of the EMF is an upper limit because the forward model at its maximum metal mass fraction configuration is less dense than the planet. All stellar ages were taken from the references listed in Table \ref{tab:nomads_stellar_prop}. }
    \begin{tabular}{lcccc}
    \hline
TOI	    & $T_{\rm{eq}}$     & $a$	  & Age    & Envelope Mass  \\
        & (K)               & (AU)    & (Gyr)  & Fraction (EMF) \\
\hline
193.01	& 1974.1  & 0.01679 & 2.0 &$0.0\substack{+0.1\\-0.00}$		 \\
576.01	& 1273.3  & 0.0641	 & 2.1 &$0.28\substack{+0.07\\-0.06}$	 \\	
465.01	& 850.82  & 0.0453	 & 6.4 &$0.18\substack{+0.02\\-0.02}$	 \\	
118.01	& 1073.4  & 0.06356 & 10.0 &$0.07\substack{+0.00\\-0.0}$		 \\
880.01	& 953.56  & 0.05092 & &$0.12\substack{+0.02\\-0.01}$	 \\	
192.01	& 973.68  & 0.0457	 & &$0.60\substack{+0.10\\-0.10}$	 \\	
2498.01	& 1443.5  & 0.0491	 & 3.6 &$0.17\substack{+0.05\\-0.05}$	 \\	
829.01	& 1207.2  & 0.04177 & &$0.00\substack{+0.03\\-0.00}$	 \\	
181.01	& 895.69  & 0.0539	 & 5.4 &$0.34\substack{+0.01\\-0.01}$	 \\	
132.01	& 1531.8  & 0.026	 & 6.34 &<0.01 \\
426.01$^{*}$	& 1796.8  & 0.02342 & &$0.0001\substack{+0.002\\-0.000}$ \\		
824.01	& 1253.8  & 0.02177 & 7.5 &<0.01 \\
969.01$^{*}$	& 1079.4  & 0.02636 & 4.8 &$0.004\substack{+0.002\\-0.000}$	 \\	
849.01	& 1966.5  & 0.01598 & 6.7 &<0.01 \\
5632.01	& 993.81  & 0.0479	 & 9.0 &$0.26\substack{+0.05\\-0.07}$	 \\	
3071.01	& 2161.4  & 0.0249	 & &$0.00\substack{+0.13\\-0.00}$	 \\	
1839.01$^{*}$	& 1537.0  & 0.02421 & &<0.0001 \\
5559.01$^{*}$	& 1482.1  & 0.0321	 & 4.9 &$0.003\substack{+0.001\\-0.001}$	 \\	
1117.01$^{*}$	& 1532.4  & 0.03304 & 4.42 &$0.001\substack{+0.002\\-0.001}$	 \\	
1853.01	& 1491.7  & 0.021	 & &<0.01 \\
544.01$^{*}$	& 1059.2  & 0.0225	 & 9.0 &<0.0001 \\
2196.01	& 1857.2  & 0.02234 & 4.5 &<0.01 \\
179.01	& 972.14  & 0.05    & 0.4 &<0.01 \\
332.01	& 1873.8  & 0.0159	 & 5.0 &<0.01 \\
1130.02	& 819.69  & 0.04457 & 4.1 &<0.01 \\
908.01	& 1348.3  & 0.04165 & 4.6 &$0.00\substack{+0.04\\-0.00}$	 \\	
5094.01$^{*}$	& 701.75  & 0.031	 & 1.6 &$0.03\substack{+0.01\\-0.01}$	 \\	
442.01	& 713.12  & 0.0417	 & &$0.07\substack{+0.03\\-0.04}$	 \\	
682.01$^{*}$	& 953.00  & 0.06977 & &$0.04\substack{+0.01\\-0.01}$	 \\	
    \hline
\end{tabular}
    \vspace{2mm}
    \begin{flushleft}
   {\bf Notes:} Targets denoted with $^{*}$ have EMFs calculated from the \cite{lopez2014understanding} models as they are too dense and not massive enough to be gas giants and are, therefore, in the sub-Neptune regime. All other EMFs have been calculated using GASTLI. 
    \end{flushleft}
    \label{tab:EMF}
\end{table}

\begin{figure}
    \centering
    \subfloat{\includegraphics[width=0.47\textwidth]{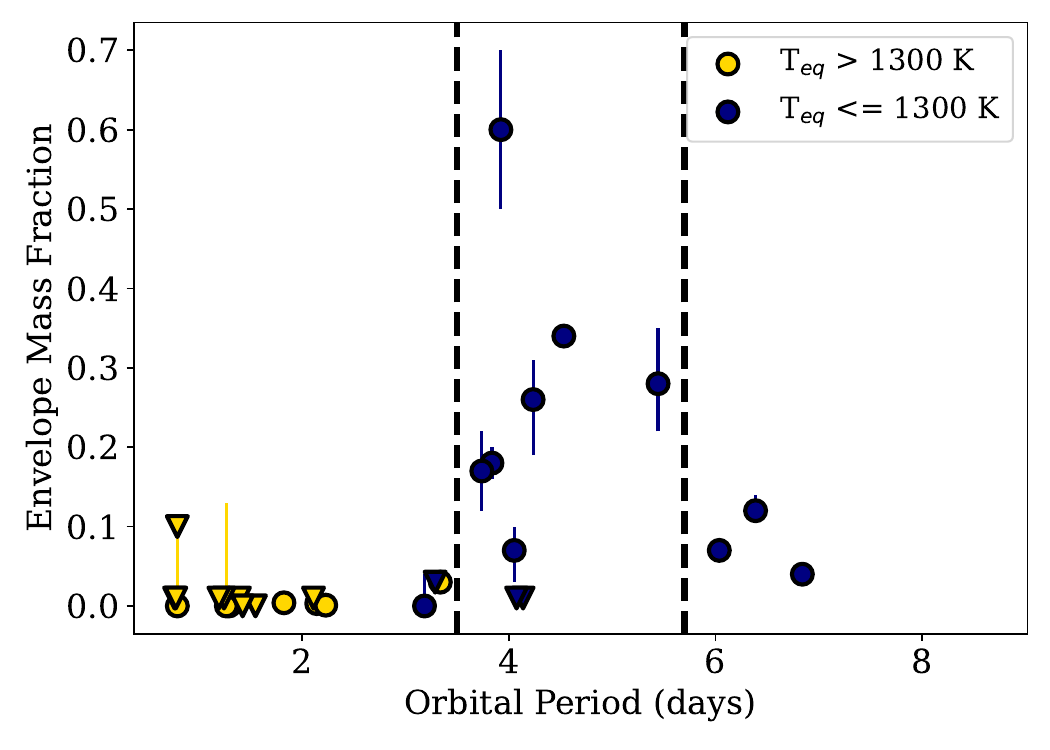}} \\ \vspace{-3mm}
    \subfloat{\includegraphics[width=0.47\textwidth]{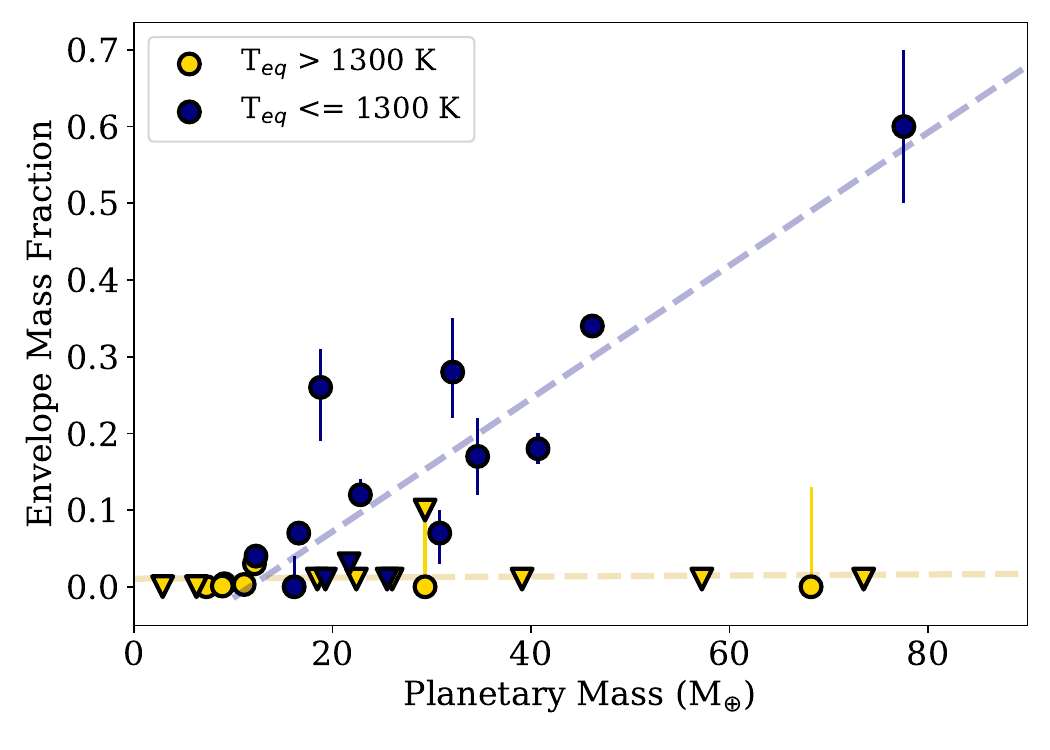}}
    \caption{{\it Top:} We show the distribution of Envelope Mass Fractions (EMF) against planetary orbital period for 29 of our `gold' sample of Neptunian desert planets. For reference, Saturn's EMF = 0.75-0.83 at M = 95.16 $M_{\oplus}$, while Neptune's EMF = 0.15-0.35 at M = 17.15 $M_{\oplus}$, depending on whether their measured atmospheric metallicities are used or a solar envelope is assumed \citep[see][for a detailed comparison with Solar System planets]{acuna2024gastli}. A dashed black line is shown at 3.5 and 5.7~days which represents the Neptune ridge defined in \citet{castro2024mapping}. {\it Bottom:} The distribution of EMF against planetary mass for 29 of our `gold' sample of Neptunian desert planets. For each temperature range a linear trend in EMF has been fitted and shown as a dashed line. In both, the sample has been split according to the planetary equilibrium temperature with T$_{\rm{eq}}$ > 1300~K in yellow and T$_{\rm{eq}} \leq$ 1300~K in blue. Triangle markers represent planets where only an upper limit on the EMF was derived.}
    \label{fig:EMF}
\end{figure}

\subsection{Results}
\label{sec:emf_results}
In Figure \ref{fig:EMF} (upper panel), we show the EMF of 29 targets within our `gold' sample, which had planetary mass and radius measurements, plotted as a function of orbital period. There is a clear split in EMF between planets with $T_{\rm{eq}}$ $\lesssim$ 1300~K and $T_{\rm{eq}}$ $\gtrsim$ 1300~K, which corresponds to $P_{\rm orb}$ $\sim$~3.5~days. Up until $\sim$3.5~days, the EMFs remain compatible with zero. For planets with orbital periods greater than $\sim$3.5 days, we find many EMFs larger than zero, which range between 5$\%$ and 60$\%$.

We studied whether the EMF-period distribution presented above could be biased by an inhomogeneous distribution of the planetary masses below and above the $\sim$3.5\,day orbital period division. In Figure~\ref{fig:EMF} (lower panel), we show EMF as a function of planetary mass. Both groups of planets are very well represented up to $\simeq$~40$M_{\oplus}$. Larger masses up to $\simeq$~80$M_{\oplus}$, while less frequent, are also found in both samples. The EMFs of the close-in group ($P_{\rm orb}$ $\lesssim$ 3.5 days) remain invariant across the wide range of masses, while the EMFs of the cooler group ($P_{\rm orb}$ $\gtrsim$ 3.5 days) increase with the planet mass. We compute the Pearson correlation for these EMFs, which yields a coefficient of 0.87 (p-value = 5.2 $\times 10^{-5}$), indicating a linear dependency between the EMF and planetary mass for planets with $T_{\rm{eq}} < 1300$~K. Therefore, we conclude that the EMF split below and above $\simeq$ 3.5 days is present in Neptunes with a wide range of masses, from $\simeq$ 20$M_{\oplus}$ to (at least) $\simeq$ 80$M_{\oplus}$. To see the EMF plotted in radius-period space and contextualized within the Neptunian landscape derived in \citet{castro2024mapping} we refer the reader to \S \ref{sec:discussion}. 

To study photo-evaporation in our `gold' sample, we can follow the framework of \cite{owen2017evaporation} to estimate the photo-evaporation mass-loss timescale, $t_{\dot{m}}$, of these planets as follows:

\begin{equation}
    t_{\dot{m}} = \frac{M_{\rm{env}}}{\dot{m}},
\end{equation}

where $M_{\rm{env}}$ is the mass of the envelope and $\dot{m}$ is the mass-loss rate of the envelope. The latter can be calculated as:

\begin{equation}
    \dot{m} = \eta \frac{\pi R_{\rm{p}}^3 L_{\rm{HE}}}{4\pi a^2 G M_{\rm{p}}},
\end{equation}

where $\eta$ = 0.1. The magnitude and evolution of the high energy flux, $L_{\rm{HE}}$, is adopted from \citep{jackson2012coronal}:

\begin{equation}
    L_{\rm{HE}} = 
    \begin{cases}
  L_{\rm{sat}} & \text{for }t < t_{\rm{sat}}\\    
  L_{\rm{sat}}\left(\frac{t}{t_{\rm{sat}}}\right)^{-1-a_0} & \text{for }t \geq t_{\rm{sat}}    ,
\end{cases}
\end{equation}

where $a_0$ = 0.5, $t_{\rm{sat}}$ = 100~Myr, $t$ is the stellar age, and \newline $L_{\rm{sat}} \approx 10^{-3.5}L_{\odot}(M_{\star}/M_{\odot})$. 

Overall, planets with $t_{\dot{m}} \gtrsim$ 100~Myr are stable to mass-loss while those with $t_{\dot{m}} \lesssim$ 100~Myr are unstable and evolve towards a lower mass atmosphere \citep[see Figure 3 in][]{owen2018photoevaporation}. From our Neptunian `gold' sample, five have $t_{\dot{m}} \lesssim$ 100~Myr and amongst these they all have EMF < 0.01 and lie outside the revised Neptune desert boundaries from \citet{castro2024mapping}. Therefore, these five targets are likely to have undergone photo-evaporation during their lifetime which has resulted in a low envelope fraction. 

The observed EMF distribution aligns very well with the revised Neptunian landscape mapped in \citet{castro2024mapping}. The authors find that, in the super-Neptune and sub-Jovian regime (i.e. 10$M_{\oplus}$ < $M_{\rm p}$ < 100$M_{\oplus}$), the desert does not extend triangularly out to orbital periods as large as 10 -- 15 days, but instead it has a constant-period vertical boundary at $\simeq$3.2 days that represents a sharp planet occurrence drop. Therefore, our results indicate that super-Neptunes and sub-Jovians outside the desert have extensive envelopes while those within the desert lack them. \citet{castro2024mapping} also find that, far from being a desert, the $\simeq$ 3.2 -- 5.7 days orbital period region has an overabundance of planets when compared to larger-period orbits, which they call the Neptunian ridge. In our analysis, there are three planets with orbital periods greater than 5.7 days, which show EMFs below 15$\%$, suggesting a drop off in this region (see Fig.~\ref{eq:cmf_correction}). This distribution suggests the existence of a sweet spot for Neptune planets to retain a large H-He envelope, between $\simeq$ 3.5 and $\simeq$ 5.7 days. We note, however, that the reduced envelopes for planets with periods larger than $\simeq$ 5.7 days needs to be further studied with a larger planet sample. A natural question that is raised here is whether the tentative EMF drop off at $\simeq$ 5.7~days reflects the transition between the ridge and the savanna found in \citet{castro2024mapping}. Additional mass and radius constraints of Neptune planets in this region are necessary to reach a firm conclusion. 

\section{Host Metallicity}
\label{sec:metallicity}
It has been well established that there is a correlation between the host star metallicity and the occurrence rate of giant exoplanets. This is known as the planet-metallicity correlation, where shortly after the first exoplanet detections around solar-type stars, short period gas giants were found to orbit more metal-rich stars \citep{gonzalez1997stellar, santos2001metal, santos2004spectroscopic, fischer2005planet}. The metal content in a protoplanetary disk is key for core accretion and relates directly to the newly forming host star \citep{pollack1996formation, alibert2004migration}. Therefore, it was determined that the main formation mechanism for giant planets in close orbits around solar-type stars was core accretion \citep{ida2004toward}. The mass budget for forming planets via core accretion is a function of metallicity and mass in the protoplanetary disk - regardless of the metallicity, if there is not sufficient mass then giant planets will not form \citep{johnson2010giant, mordasini2012extrasolar, ghezzi2018retired, manara2018protoplanetary, mulders2021mass}. There is a strong trend of increasing occurrence of Jupiter and sub-Saturn planets with host star metallicity \citep{fischer2005planet, sousa2011spectroscopic}. For stellar hosts of F,G,K-type there is a break at 4 - 10~M$_{\rm{J}}$ when the occurrence no longer depends on metallicity, suggesting a differing formation mechanism \citep{santos2017observational, narang2018properties, schlaufman2018evidence, maldonado2019connecting}. Close-in Neptunes (P$_{\rm{orb}}$ < 10~days) are also found around more metal-rich stars, but their occurrence is not as strongly linked to metallicity as hot Jupiters \citep{petigura2018california, osborn2020investigating, wilson2022influence}. 

\begin{figure}
    \centering
    \includegraphics[width=0.47\textwidth]{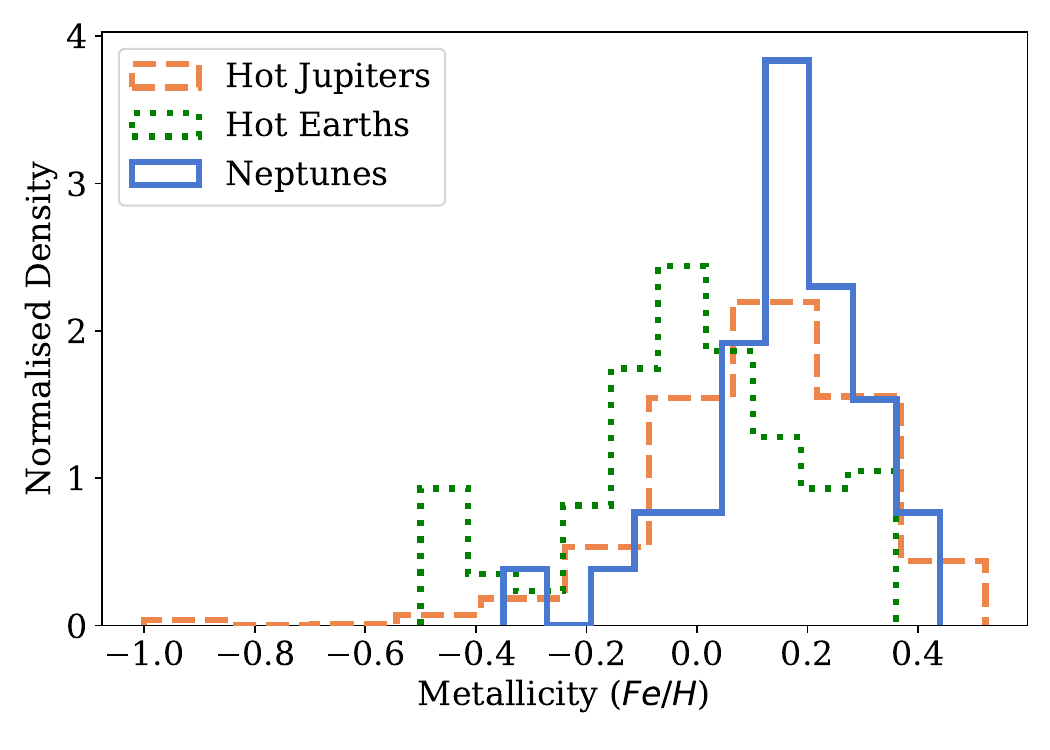}
    \caption{The metallicity distribution Fe/H for our `gold' sample of Neptune desert targets (blue-solid) compared to the distribution of hot Jupiters (orange-dashed) and hot Earths (green-dotted). The hot Jupiter and hot Earth samples were taken from the TEPCat catalogue (\url{https://www.astro.keele.ac.uk/jkt/tepcat/}) both with P$_{\rm{orb}} \leq 10$~days and M$_{\rm{p}}$ between 0.1 - 13~M$_{\rm{J}}$ and M$_{\rm{p}} \leq$ 2~M$_{\oplus}$, respectively.}
    \label{fig:metallicity}
\end{figure}

\citet{dai2021tks} showed that the metallicity of ultra-short-period terrestrial planets follows a solar-like distribution which is similar to that of longer-period sub-Neptune planets \citep{winn2017absence}. Within their sample, they had three hot Neptune targets which all orbit metal-rich stars, indicating an enhanced metallicity of hot Neptune host stars. Previous work by \citet{dong2018lamost} also showed a preference for hot Neptunes with P$_{\rm{orb}} \leq $10~days to occur around metal-rich stars. Recently, \citet{vissapragada2024hottest} demonstrated with a larger sample that planets in the Neptune desert and the ridge orbit host stars with enhanced metallicity. In Figure \ref{fig:metallicity}, we show the distribution of host star metallicities (using Fe/H) for our sample of `gold' Neptune desert targets compared to the distribution for hot Jupiters and hot Earths drawn from the TEPCat catalogue \citep{southworth2011homogeneous}. Overall, we find an enhanced metallicity compared to solar for our sample of 33 Neptune desert planets which aligns with previous results despite the differing samples. The two-sample Kolmogorov-Smirnov (KS) test, comparing the hot Jupiter and Neptune desert samples, yields KS = 0.20 and p-value = 0.13, where the p-value is above the threshold of 0.05, implying that the distributions are indistinguishable. On the other hand, when comparing the hot Earths and Neptune desert samples, this yields KS = 0.51 and p-value = 2.1 $\times 10^{-6}$, where the p-value is below the 0.05 threshold, implying the distributions are drawn from two different samples. 

\section{Discussion}
\label{sec:discussion}
\begin{figure}
    \centering
    \includegraphics[width=0.47\textwidth]{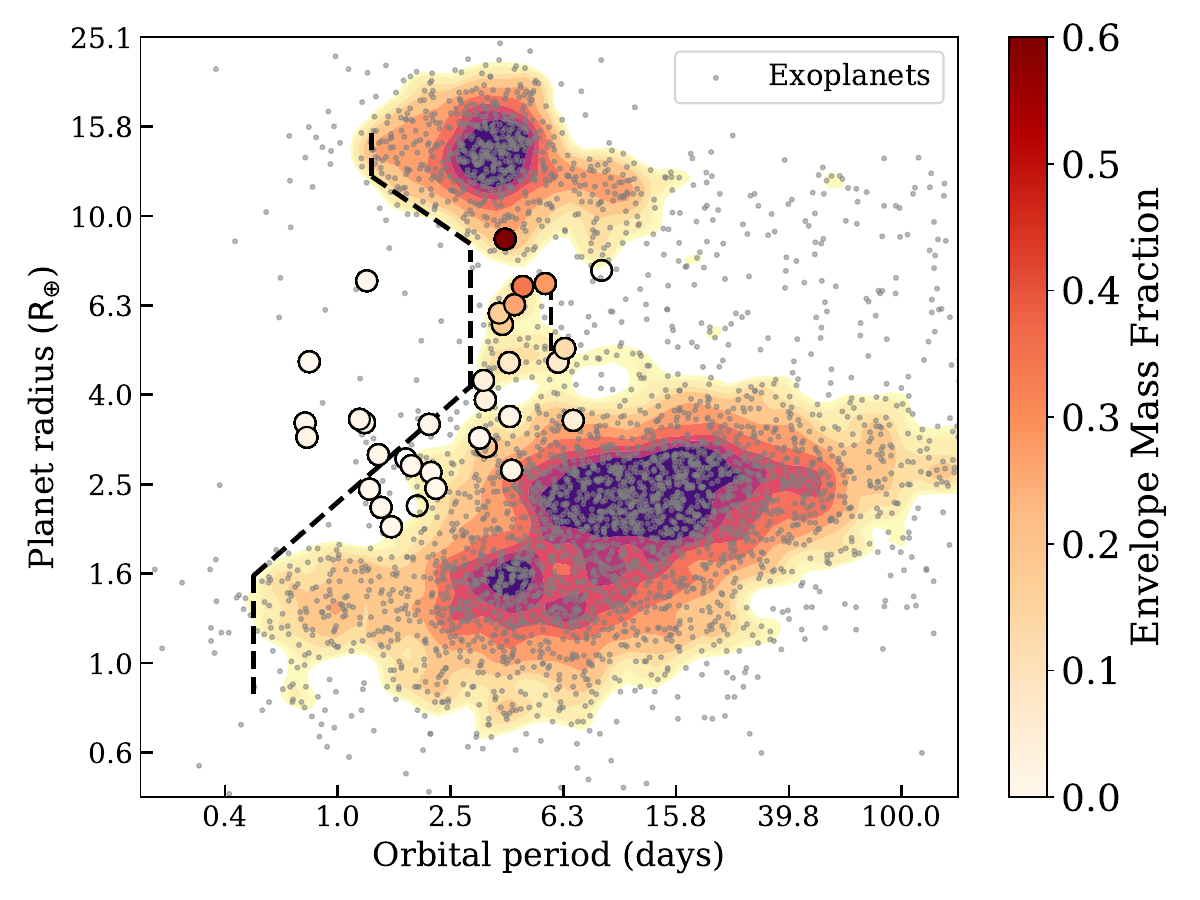}
    \caption{The period-radius diagram of close-in exoplanets highlighting the `gold' Neptunian desert sample from this work. The sample has been colour coded according to the envelope mass fractions determined in \S \ref{sec:emf} where transparent points represent those where no EMF was determined due to a lack of known planetary mass (see \S \ref{sec:gold_sample}). We highlight the population based boundaries of the Neptunian desert, ridge, and savanna as derived in \citet{castro2024mapping}. The background exoplanet sample in grey was collected from the NASA Exoplanet Archive (\url{https://www.exoplanetarchive.ipac.caltech.edu}, accessed on 28 January 2025) and the contour represent the density of this sample. This plot was generated using the {\tt nep-des} Python code (\url{https://github.com/castro-gzlz/nep-des}).}
    \label{fig:nep_desert_dis}
\end{figure}

In our envelope mass fraction analysis for our `gold' sample of Neptune targets we found a trend in EMF with T$_{\rm{eq}}$ < 1300~K, where the EMF increases linearly for higher planetary mass. This trend has also been found in gas giant planets and is attributed to core accretion \citep{thorngren2016mass}. The presence of large H/He envelopes in massive planets indicates that they not only managed to accrete gas before disk dispersion, but conserved a significant amount of gas despite the irradiation of the star \citep{rogers2015most}. In contrast, at higher equilibrium temperatures, of T$_{\rm{eq}}$ > 1300~K, the EMFs remain largely compatible with zero, regardless of the planetary mass. This does not necessarily mean that these planets did not undergo core accretion, but their pathway could be accretion of a massive envelope followed by atmospheric escape, which could have erased the mass-EMF trend at high equilibrium temperatures and short orbital periods. Interestingly, the EMF split at T$_{\rm{eq}}$ = 1300~K corresponds to an orbital period of $\sim$3.5~days, which coincides with the transition between the Neptunian desert and the recently identified Neptunian ridge at $\sim$3.2~days \citep{castro2024mapping}. In Fig.~\ref{fig:nep_desert_dis}, we plot the period-radius distribution of our `gold' sample colour-coded according to their inferred EMFs, where we can appreciate the Neptunes with the larger EMFs clustering in the upper region of the ridge near the hot Jupiter pileup. 

The lack of an envelope in a close-in planet suggests that the planet's proximity to its host star may have caused significant atmospheric escape. XUV-photoevaporation \citep[e.g.][]{rogers2021photoevaporation} and core-powered mass loss \citep[e.g.][mainly relevant for sub-Neptunes]{gupta2019sculpting} are both efficient mechanisms in removing the envelope of exoplanets. XUV-photoevaporation is driven by the irradiation received from the star, whereas core-powered mass loss is dependent on the internal heat released by the planet's core. Given that we observe a trend of EMF with proximity to its host star in our `gold' sample, XUV-photoevaporation is favoured as the dominant atmospheric escape mechanism that shapes the composition of exoplanets in the Neptune desert. However, core-powered mass-loss could play a significant role in combination with XUV-photoevaporation. If this is the case, then the EMF should also decrease with internal heat \citep[e.g.][]{leconte2010tidal, sing2024warm, welbanks2024high}, however, this is something we cannot explore with our current data. The EMFs of our `gold' sample can provide insights into the formation and evolution histories of close-in Neptunes, and shed some light on the lack of the closer-in ones. Amongst our `gold' sample of Neptunian desert targets, there is a large range of masses (2.9~M$_{\oplus}$ -- 77.5~M$_{\oplus}$), which yields a diversity within the EMFs. Overall, this split in EMF at T$_{\rm{eq}}$ = 1300~K hints at differing formation and evolution pathways for these two populations. Moving forward, an improved precision in ages with upcoming stellar characterisation methods and missions such as PLATO \citep[for FGK stars,][]{rauer2010plato} along with eccentricities and internal heat measurements will shed light on if high internal heat can be maintained in time by tidal heating. 

We also found our `gold' sample of Neptune desert planets had enhanced stellar metallicity compared to solar. \citet{owen2018metallicity} investigated trends in planetary properties with host star metallicity for the California-Kepler survey. They found short and long period sub-Neptunes have larger radii around higher metallicity stars. They conclude the formation of sub-Neptunes around higher metallicity stars results in larger, more massive solid cores. For short orbital periods where atmospheres should be mostly stripped as a result of photo-evaporation, all small planets are larger around high metallicity stars. On the other hand, increasing the host star metallicity may result in metal-rich gaseous envelopes which can resist photo-evaporation at shorter orbital periods \citep[see][and references therein]{teske2024star}. With the enhanced metallicity of close-in Neptune planets, this could suggest their H/He envelopes may remain intact despite their proximity to the host star due to the higher metallicity conditions during planetary formation. Therefore, testing this atmospheric characterisation (i.e. through transmission, emission or phase curves) is needed to determine (i) the presence of an atmosphere and (ii) the composition. Overall, high metallicity envelopes are more likely to be retained than solar envelopes due to the high mean molecular weight. 

The two main processes which are believed to drive the distribution of close-in Neptune planets are migration and evaporation. Neptunes in the ridge (P$_{\rm{orb}}$ between $\sim$3.2 and $\sim$5.7~days) have been observed to have eccentric and polar orbits \citep[e.g.][]{correia2020warm, bourrier2023dream}, which suggests high-eccentricity migration (HEM) as the main driver in their evolution. Amongst our `gold' sample, 12 planets have P$_{\rm{orb}}$ between 3.2 and 5.7~days, of which four have eccentricities greater than 0.1 with the remaining being in close to circular/circular orbits. Additionally, there are five planets amongst these 13 which have sky-projected spin-orbit angle ($\lambda$) measurements: TOI-576.01 \citep[WASP-166~b:][$\lambda = -15.5 \pm 2.9^{\circ}$]{doyle2022wasp166}; TOI-465.01 (WASP-156~b: confirmed aligned, priv. communication); TOI-5632.01 (HAT-P-26~b: presumed aligned, priv. communication); TOI-179.01 \citep[HD 18599~b:][$\lambda = 24.0 \pm 6.7^{\circ}$]{desidera2023toi179}; and TOI-442.01 \citep[LP 714-47~b:][$\lambda = -23.9 \pm 0.8^{\circ}$]{dreizler2020toi442}. Overall, the sky-projected spin-orbit angles for these five systems are consistent with being aligned, with eccentricities close to zero, except for TOI-5632.01 and TOI-179.01 which have eccentricities less than 0.2. However, it has been shown in \citet{attia2023dream2} that there is an observational bias yielding low values for $\lambda$ when the 3D spin-orbit angle ($\psi$) is low or moderate (i.e. a value of $\lambda$ close to zero is unlikely to correspond to a $\psi$ close to zero). Therefore, for these five with measured sky-projected spin-orbit angles, we need to retrieve the 3D spin-orbit angle to assess the full architecture and misalignment of the systems. 

For giant planets, an association between obliquity and eccentricity is expected, which is a result of HEM \citep[][]{dawson2018origins}. The process of HEM involves a proto-planet being launched into an eccentric orbit caused by planet-planet scattering \citep{beauge2012multiple}, von Zeipel-Lidov-Kozai oscillations \citep{fabrycky2007shrinking}, or secular chaos \citep{wu2011secular}. Therefore, the lack of planets with both high spin-orbit angles and eccentric orbits amongst our planets with P$_{\rm{orb}}$ between $\sim$3.2 and $\sim$5.7~days could suggest HEM is not at play here. However, the age of the system must be taken into consideration as the orbit will shrink and circularise over time due to tidal effects, which is a much slower process compared to the alignment of the stellar spin axis \citep{hut1981tidal}. With this in mind, HEM could be dominating planet migration across the entire Neptune desert and ridge, where planets detected in the ridge have recently migrated and as a result have not had time to circularise their orbits and align their stellar spin axis \citep[see further discussions by][]{bourrier2018orbital, attia2021jade,castro2024mapping}. As a consequence of the shrinking and circularisation of the planetary orbits, photo-evaporation takes over and planets lose their gaseous envelopes leading them to be within the Neptune desert. 

Amongst our `gold' sample, 20 planets have P$_{\rm{orb}} \leq$ 3.5~days of which five have eccentricities greater than 0.1, with the remaining being in circular orbits. There is only one target which has an obliquity measurement and that is TOI-5094.01 (GJ 3470~b) which has a polar orbit with $\lambda = 101\substack{+29 \\ -14}^{\circ}$ \citep{stefansson2022toi5094}. In \citet{stefansson2022toi5094}, they conclude the formation scenario of TOI-5094.01 likely involved multi-body dynamical interactions, which includes interactions with a massive distant perturber. Overall, more obliquity measurements of Neptune desert and Neptune ridge planets are needed to aid in understanding how these planets arrive at their observed orbits and what their dynamic histories entailed. This is something which is actively being worked on through the ATREIDES programme (ESO program ID 112.25BG, P.I. Bourrier: Bourrier et al. in prep).  

\section{Conclusions}
In this paper, we have carried out a homogenous analysis of planets within the Neptune desert and its surroundings, for which there are precise mass and radius measurements. Our target sample was created from all known TOIs cutting on magnitude, declination, stellar host radius, stellar host effective temperature, planetary orbital period, and {\tess} follow-up (TFOP) status. We then further refined our sample using a merit ranking based on visual magnitude and location within the original Neptunian desert \citep{mazeh2016dearth}. After making a cut on merit ranking, a final full sample of 64 targets was identified, and through further HARPS observations we ensured that each target in the sample had precise radial velocity measurements. After observations, the sample consisted of 46 confirmed planets and 18 null detections with upper limits on the candidate mass. We also focused on Neptune planets specifically by identifying a `gold' sample with planetary radii between 2 - 10~R$_{\oplus}$. The results of our analysis are summarised here:

\begin{itemize}
    \item Cross-matching our sample with Gaia DR3 we found all targets in our full sample with upper mass limits (i.e. non-detection) have a preference to lie higher on the main sequence where stellar hosts are more likely to be members of binary and triple systems. 
    \item 23\% of planets in our full sample are in binary systems with a spread across the close-in period-radius space.
    \item 26\% of planets in our full sample are in multi-planet systems, which are clustered near the lower boundary of the Neptune desert. 
    \item Two clusters appear in the mass-radius diagram of our full sample of Neptunian targets with a split in orbital period at $\sim$3~days, similar to that seen in the EMF analysis. 
    \item We found a bimodal distribution of EMFs at equilibrium temperatures of 1300~K (i.e. orbital periods of $\sim$3.5 days) for our `gold' sample, which aligns well with the transition between the Neptunian desert and the recently identified Neptunian ridge at $\sim$3.2 days. 
    \item Planets with T$_{\rm{eq}}$ > 1300~K ($P_{\rm orb}$ $\lesssim$ 3.5 days) have very low to zero EMF, suggesting photo-evaporation is a key evolution mechanism in the Neptunian desert.
    \item For planets with T$_{\rm{eq}}$ < 1300~K ($P_{\rm orb}$ $\gtrsim$ 3.5 days) there is a trend in EMF which increases for higher planetary mass, as predicted for gas giant planets formed via core accretion.
    \item We found an enhanced metallicity when compared to solar for our `gold' sample of close-in Neptunes which matches the distribution for hot Jupiters, in line with previous results. 
\end{itemize}

Overall this paper presents a well-defined sample of desert planets with transit and radial velocity measurements, allowing a unified study encompassing planet radius, host stars, orbits, mass, densities, and interior structure probed through the EMF. With this work we provide robust experimental constraints to inform theories of how planet in and around the Neptune desert form, evolve, and persist in this hostile region of parameter space.

\section*{Acknowledgements}
This research has made use of the NASA Exoplanet Archive, which is operated by the California Institute of Technology, under contract with the National Aeronautics and Space Administration under the Exoplanet Exploration Program. This research has made use of the Exoplanet Follow-up Observation Program (ExoFOP; DOI: 10.26134/ExoFOP5) website, which is operated by the California Institute of Technology, under contract with the National Aeronautics and Space Administration under the Exoplanet Exploration Program.

We include data in this paper collected by the {\tess} mission, where funding for the {\tess} mission is provided by the NASA Explorer Program. This work presents results from the European Space Agency (ESA) space mission Gaia. Gaia data are being processed by the Gaia Data Processing and Analysis Consortium (DPAC). Funding for the DPAC is provided by national institutions, in particular the institutions participating in the Gaia MultiLateral Agreement (MLA). The Gaia mission website is \url{https://www.cosmos.esa.int/gaia}. The Gaia archive website is \url{https://archives.esac.esa.int/gaia}.

This research was funded in whole or in part by the UKRI, (Grants ST/X001121/1, EP/X027562/1). This work has been carried out in the frame of the National Centre for Competence in Research PlanetS supported by the Swiss National Science Foundation (SNSF). This project has received funding from the European Research Council (ERC) under the European Union's Horizon 2020 research and innovation programme (project {\sc Spice Dune}, grant agreement No 947634). A.C.-G. is funded by the Spanish Ministry of Science through MCIN/AEI/10.13039/501100011033 grant PID2019-107061GB-C61. S.C.C.B acknowledges the support from Funda{\c c}{\~ a}o para a Ci{\^ e}ncia e Tecnologia (FCT) in the form of work contract through the Scientific Employment Incentive program with reference 2023.06687.CEECIND. O.D. also acknowledges FCT - through national funds by the following grants: UIDB/04434/2020; UIDP/04434/2020; 2022.06962.PTDC. X.D acknowledges the support from the European Research Council (ERC) under the European Union’s Horizon 2020 research and innovation programme (grant agreement SCORE No 851555) and from the Swiss National Science Foundation under the grant SPECTRE (No 200021\_215200).  J.S.J. greatfully acknowledges support by FONDECYT grant 1240738 and from the ANID BASAL project FB210003. J.L.-B. is funded by the Spanish Ministry of Science and Universities (MICIU/AEI/10.13039/501100011033) and NextGenerationEU/PRTR grants PID2019-107061GB-C61, CNS2023-144309, and PID2023-150468NB-I00. L.P acknowledges the support from the framework of the NCCR PlanetS supported by the Swiss National Science Foundation under grants 51NF40\_205606. J.R. and N.C.S. acknowledge funding by the European Union (ERC, FIERCE, 101052347). A.S. is supported by the "Programme National de Plan{\' e}tologie" (PNP) of CNRS/INSU co-funded by CNES.

Views and opinions expressed are however those of the author(s) only and do not necessarily reflect those of the European Union or the European Research Council. Neither the European Union nor the granting authority can be held responsible for them.

For the purpose of open access, the author has applied a Creative Commons Attribution (CC BY) licence (where permitted by UKRI, ‘Open Government Licence’ or ‘Creative Commons Attribution No-derivatives (CC BY-ND) licence’ may be stated instead) to any Author Accepted Manuscript version arising from this submission.

\section*{Data Availability}
All {\tess} SPOC data are available from the NASA MAST portal. Gaia DR2 and DR3 data is available from the Gaia Archive. The full HARPS RV data products are publicly available from the ESO archive. The full list of all 64 targets and their properties is available as supplementary material online for the community. 



\bibliographystyle{mnras}
\bibliography{NOMADS}




\appendix

\section{Stellar Parameter Fitting}
\label{sec:stellar_prop}
The stellar spectroscopic parameters ($T_{\mathrm{eff}}$, $\log g$, microturbulence, [Fe/H]) were estimated using the ARES+MOOG methodology. The methodology is described in detail in \citet[][]{Sousa-21, Sousa-14, Santos-13}. We used the latest version of ARES \footnote{The last version, ARES v2, can be downloaded at https://github.com/sousasag/ARES} \citep{Sousa-07, Sousa-15} to consistently measure the equivalent widths (EW) on the list of iron lines. The list of lines used in this analysis is the one presented in \citet[][]{Sousa-08}; however, for cooler stars we used the list of lines presented in \citet[][]{Tsantaki-2013}, which is more appropriate for stars with temperature below 5200~K. For the stars that were spectroscopy analysed in this work we first compiled all the individual HARPS observations to produce a higher SNR combined spectrum. In the analysis, we use a minimization process to find the ionization and excitation equilibrium to converge for the best set of spectroscopic parameters. This process makes use of a grid of Kurucz model atmospheres \citep{Kurucz-93} and the radiative transfer code MOOG \citep{Sneden-73}. We also derived a more accurate trigonometric surface gravity using recent {\sl Gaia} data following the same procedure as described in \citet[][]{Sousa-21}. In this process, we also estimate stellar masses and stellar radii by using the calibrations presented in \citep[][]{Torres-2010}. All results from this work can be found in Table \ref{tab:nomads_stellar_prop}.

\begin{table*}
    \centering
    \caption{The Stellar properties for all targets in our Neptunian desert sample. From left to right the columns represent the TESS Object of Interest (TOI) ID, the TESS Input Catalogue (TIC) ID, the Gaia DR3 ID from \citet{vallenari2023gaiadr3}, Right Ascension, Declination, stellar mass, stellar radius, metallicity, stellar effective temperature, and the literature references for all stellar parameters within this table. The metallicity for host stars where all other information was from the TIC Catalogue was taken from EXOFOP (\url{https://www.ExoFOP.ipac.caltech.edu/tess/}) where spectroscopic observations from CHIRON, TRES or FIES were used. This full table is available as supplementary material online.}
    \label{tab:nomads_stellar_prop}
    \resizebox{0.99\textwidth}{!}{    
    \begin{tabular}{lllccccccr}
    \hline
    TOI	   &  TIC ID    &  Gaia DR3 ID            &  RA          &  Dec        &  $M_{\star}$  &  $R_{\star}$    & $[Fe/H]$ & $T_{\rm{eff}}$ &  Reference \\ 
           &   	     &                         &              &   	        &  (M$_{\odot}$) &  (R$_{\odot}$) & (dex)  & (K) 	        &            \\ 
    \hline
    422    &  117979455	&  2980377518655206784    &  71.8027847	 &  -17.2533769	& 1.04	& 1.45  & -0.198 	& 5894   & 	This Work   \\
    193    &  183985250	&  2307504062045842688	  &  358.6689201 &  -37.6282335	& 1.02	& 0.94  &  0.27   	& 5443   & 	\citet{jenkins2020toi193}   \\
    1967   &  320079492	&  5236556416614488576    &  174.8449836 &  -65.1762954	& 1.01	& 1.49  & -0.24	    & 5699   & 	TIC Catalogue   \\
    855    &  269558487	&  2532584370108897920    &  15.5430449	 &  -2.1445754	& 1.04	& 1.40  & -0.484	& 6177   & 	This Work   \\
    728    &  96097215	&  3819928558555030528    &  146.8917330 &  -7.3307350	& 0.94	& 0.97  & -0.077	& 5790   & 	This Work   \\
    355    &  183593642	&  5013703860801457280    &  17.8882142	 &  -36.7091202	& 1.15	& 1.49  & -0.031	& 6053   & 	This Work   \\
    576    &  408310006	&  5664957444179338240	  &  144.8750965 &  -20.9824175	& 1.19	& 1.22  & 0.19	    & 6050   & \citet{hellier2019toi576}   \\
    465    &  270380593	&  2514360548993901312	  &  32.7820953	 &  2.4181993	& 0.842	& 0.76  & 0.24	    & 4910   & 	\citet{demangeon2018toi465}   \\
    1975   &  467281353	&  5337617753004196864    &  168.0760684 &  -60.9325783	& 1.02	& 1.36  & -0.068	& 5741   & 	This Work   \\
    118    &  266980320	&  6492940453524576128	  &  349.5593464 &  -56.9039886	& 0.92	& 1.03  & 0.04	    & 5527   & 	\citet{esposito2019toi118}   \\
    880    &  34077285	&  2993658970584498048	  &  94.1644735	 &  -13.9873337	& 0.81	& 0.81  & 0.14 		& 4935   & 	TIC Catalogue   \\
    271    &  259511357	&  4784439709132505344    &  70.7755204	 &  -50.2194708	& 1.15	& 1.33  & -0.01	    & 6114   & 	TIC Catalogue   \\
    851    &  40083958	&  2426349694072115968 	  &  7.3878627	 &  -10.4682720	& 0.961	& 0.77  & -0.32	    & 5485   & 	TIC Catalogue   \\
    141    &  403224672	&  6407428994690988928	  &  338.9836867 &  -59.8648408	& 1.068	& 1.10  & -0.04	    & 5978   & 	\citet{espinoza2020toi141}   \\
    192    &  183537452	&  6534414719318886144	  &  357.8791874 &  -39.9071339	& 0.825	& 0.80  &  0.11  	& 4800   & \citet{hellier2010toi192} \\
    510    &  238086647	&  5508330367833229824 	  &  104.9590024 &  -49.5074900	& 1.07	& 1.29  & -0.48	    & 5883   & 	TIC Catalogue   \\
    2498   &  263179590	&  3330907293088717824	  &  95.4162203	 &  11.2516449	& 1.12	& 1.26  & 0.167	    & 5905   & \citet{frame2023toi2498}\\
    829    &  276128561	&  199033466340798464	  &  224.2083795 &  -37.0429856	& 0.9	& 0.94  & 0.23 	    & 5250   & 	TIC Catalogue   \\
    2358   &  124095888	&  6180723574082338560 	  &  196.2858333 &  -31.9853802	& 1.20	& 1.20  & 0.089	    & 6354   & 	This Work   \\
    641    &  49079670	&  2925175888849429888 	  &  101.0716957 &  -23.8191840	& 1.002	& 1.04  & 0.12	    & 5646   & 	TIC Catalogue   \\
    4524   &  333657795	&  30398648945512960	  &  49.1781401	 &  15.6563261	& 0.90	& 1.13  & -0.07	    & 5596   & \citet{murgas2022toi4524}\\
    431    &  31374837	&  2908664557091200768	  &  83.2692529	 &  -26.7238488	& 0.78	& 0.73  &  0.2	    & 4850   & 	\citet{osborn2021toi431}   \\
    564    &  1003831	&  5710154317045164416	  &  130.2951531 &  -16.0363289	& 0.998	& 1.08  &  0.143	& 5640   & 	\citet{davis2020toi564}\\
    4461   &  149282072	&  4511764257124805248 	  &  279.1468853 &  16.4524409	& 0.76	& 0.93  & -0.030	& 4966   & 	This Work   \\
    2673   &  1018843	&  5646258122946138496	  &  130.9321761 &  -26.9808256	& 0.99	& 1.24  &  0.328	& 5601   &  TIC Catalogue   \\
    1943   &  382980571	&  6054233458679730432	  &  183.7040893 &  -63.0867209	& 1.03	& 0.95  & -0.27     & 5742   & 	TIC Catalogue   \\
    5556   &  55315929	&  606477252238780160	  &  140.3390797 &  14.3678672	& 1.18	& 1.22  & 0.26	    & 5896   & 	\citet{frustagli2020toi5556} \\
    181    &  76923707	&  6552346035980010624	  &  352.1725597 &  -34.4914103	& 0.822	& 0.74  & 0.27     & 4994   & 	\citet{mistry2023toi181}\\
    261    &  63898957	&  2345033662373296768	  &  15.2181578	 &  -24.4239950	& 1.069	& 1.28  &  -0.35	& 5890   & 	\citet{hord2024toi261}  \\
    132    &  89020549	&  6520880040423258240	  &  338.3996691 &  -43.4368794	& 0.97	& 0.90  & 0.16	    & 5397   & \citet{diaz2020toi132}\\
    426    &  189013224	&  2983316311375470976	  &  79.1021940	 &  -15.5102453	& 1.023	& 0.98  & 0.10   	& 5731   & 	TIC Catalogue   \\
    824    &  193641523	&  5880886001621564928	  &  222.1654570 &  -57.5888878	& 0.710	& 0.69  &  -0.092	& 4600   & 	\citet{burt2020toi824}\\
    969    &  280437559	&  3087206553745290624	  &  115.1366811 &  2.0985771	& 0.734	& 0.67  &  0.175	& 4435   & \citet{lillo2023toi969}\\
    849    &  33595516	&  5023809953208388352	  &  28.7158074	 &  -29.4217057	& 0.929	& 0.91  &  0.19	    & 5375   & \citet{armstrong2020rtoi849}\\
    4537   &  251039147	&  2623358710067927680 	  &  340.5915818 &  -6.8710477	& 1.09	& 1.11  & 0.159	    & 5992   & 	This Work   \\
    5632   &  420779000	&  3668036348641580288	  &  213.1565561 &  4.0593976	& 0.816	& 0.78  &  -0.04	& 5079   & \citet{hartman2011toi5632}\\
    561    &  377064495	&  3850421005290172416	  &  148.1851364 &  6.2160946	& 0.806	& 0.84  &  -0.40	& 5372   & \citet{lacedelli2022toi561}   \\
    2427   &  142937186	&  5055663973297050624	  &  52.2917560	 &  -31.3628607	& 0.64	& 0.65  &  -0.39    & 4072   & \citet{giacalone2022toi2427}\\
    745    &  444842193	&  5307756013606927616 	  &  147.7574239 &  -55.3182994	& 1.07	& 0.97  &  -0.17	& 5881   & 	TIC Catalogue   \\
    2365   &  344085117	&  5505199886430101888 	  &  108.0174503 &  -50.2477805	& 1.08	& 1.09  & 0.211	    & 5885   & 	This Work   \\
    5005   &  282485660	&  5984530842395365248 	  &  238.1081924 &  -48.1451034	& 0.97	& 0.93  & 0.15	    & 5749   & \citet{castro2024toi5005}\\
    3071   &  452006073	&  5342880462319699200	  &  173.2790779 &  -56.5034088	& 1.29	& 1.31  & 0.35	    & 6177   & \citet{hacker2024toi3071}\\
    1839   &  381714186	&  3729117900352224768	  &  196.8331088 &  5.8520373	& 0.934	& 0.84  & 0.16   	& 5382   &  TIC Catalogue   \\
    5559   &  456945304	&  144150720342734208	  &  67.8205252	 &  19.8316864	& 0.962	& 0.94  &  0.08	    & 5662   & \citet{sozzetti2024toi5559}\\
    731    &  34068865	&  5412250540681250560	  &  146.1213719 &  -45.7790974	& 0.454	& 0.45  &  -0.01	& 3522   & \citet{lam2021toi731}\\
    1117   &  295541511	&  6436995652638923392	  &  273.6020552 &  -66.4199796	& 0.97	& 1.05  & 0.136	    & 5635   & 	Lockley et al. (submitted) \\
    1853   &  73540072	&  1233227944114002688	  &  211.4593402 &  16.9923686	& 0.837	& 0.80  &  0.11     & 4985   & \citet{naponiello2023toi1853}\\
    544    &  50618703	&  3220926542276901888	  &  82.2900764	 &  -0.3429091	& 0.630	& 0.62  &  -0.17	& 4169   & \citet{osborne2024toi544}\\
    499    &  123702439	&  5535106156033385472 	  &  115.6386899 &  -43.5534834	& 0.92	& 1.12  & -0.424	& 5922   & 	This Work   \\
    2196   &  372172128	&  6375983988633147392	  &  312.3400813 &  -70.4850751	& 1.032	& 1.04  &  0.14	    & 5634   & \citet{persson2022toi2196}\\
    179    &  207141131	&  4728513943538448512	  &  44.2620054	 &  -56.1918623	& 0.863	& 0.76  &  0.0	    & 5145   & \citet{desidera2023toi179}\\
    332    &  139285832	&  6529471108882243840	  &  348.0589587 &  -44.8764905	& 0.88	& 0.87  & 0.256	    & 5251   & \citet{osborn2023toi332}\\
    500    &  134200185	&  5509620021956148736	  &  106.5591260 &  -47.5878359	& 0.740	& 0.67  &  0.12	    & 4440   & \citet{serrano2022toi500}\\ 
    2350   &  47601197	&  2937147543250978944	  &  97.4081211	 &  -21.9943867	& 0.96	& 1.08  & 0.35	    & 5484   & 	TIC Catalogue   \\
    238    &  9006668	&  2405081733281425280	  &  349.2310737 &  -18.6066464	& 0.790	& 0.73  &  -0.114	& 5059   & \citet{mascareno2024toi238}\\
    4517   &  301289516	&  2643842302456085888	  &  351.7718285 &  -1.2853135	& 0.606	& 0.60  &  -0.26	& 4305   & \citet{bonomo2023toi4517}   \\
    1130   &  254113311	&  6715688452614516736	  &  286.3760060 &  -41.4376437	& 0.71	& 0.68  & 0.30	    & 4350   & \citet{korth2023toi1130}\\
    \hline 
    \end{tabular}}
    \vspace{-1mm}
     \begin{flushleft}
   {\bf Notes:}  For targets with parameters from the TIC it can be directly accessed through the Mikulski Archive for Space Telescopes (MAST) at \url{https://www.archive.stsci.edu/missions-and-data/tess}. A description of derived parameters from this paper can be found in Appendix \ref{sec:stellar_prop}. 
    \end{flushleft}
\end{table*}

\begin{table*}
    \centering
    \contcaption{The Stellar properties for all targets in our Neptunian desert sample.}
    \resizebox{0.95\textwidth}{!}{    
    \begin{tabular}{lllccccccr}
    \hline
    TOI	   &  TIC ID    &  Gaia DR3 ID            &  RA          &  Dec        &  $M_{\star}$  &  $R_{\star}$    & $[Fe/H]$ & $T_{\rm{eff}}$ &  Reference \\ 
           &   	     &                         &              &   	        &  (M$_{\odot}$) &  (R$_{\odot}$) & (dex)  & (K) 	        &            \\ 
    \hline
    908    &  350153977	&  4619238087059206784	  &  53.1594324	 &  -81.2507451	& 0.950	& 1.02  &  0.08	    & 5626   & \citet{hawthorn2023toi908}\\
    5094   &  19028197	&  654687847820642560	  &  119.7734753 &  15.3912012	& 0.527	& 0.50  &  0.18	    & 3622   & \citet{stefansson2022toi5094} \\
    2411   &  10837041	&  2471334872292963456	  &  20.9224402	 &  -8.7018076	& 0.64	& 0.72  &  -0.19    & 4099   & 	TIC Catalogue   \\
    442    &  70899085	&  3189306030970782208	  &  64.1902283	 &  -12.0848917	& 0.59	& 0.58  &  0.41	    & 3950   & \citet{dreizler2020toi442}\\
    682    &  429304876	&  3538925100535948416	  &  167.8854163 &  -22.0727877	& 0.969	& 0.96  &  0.44	    & 5309   & \citet{quinn2021toi682}\\
    2227   &  405425498	&  6385929006882225024    &  337.5177553 &  -67.8501425	& 1.12	& 1.15	&  0.220	& 5979   & 	This Work    \\    
    \hline 
    \end{tabular}}
    \vspace{-1mm}
    \begin{flushleft}
   {\bf Notes:}  For targets with parameters from the TIC it can be directly accessed through the Mikulski Archive for Space Telescopes (MAST) at \url{https://www.archive.stsci.edu/missions-and-data/tess}. A description of derived parameters from this paper can be found in Appendix \ref{sec:stellar_prop}. 
    \end{flushleft}
\end{table*}

\section{Upper Mass Limits on Null Detections}
\label{sec:null_results}
In the course of completing the planetary sample used in this work, we took HARPS observations of candidate targets which did not already have precise radial velocity measurements (ESO program ID 108.21YY, P.I Armstrong). Confirmed planetary results from that survey have been published separately, and Table \ref{tab:planet_prop} gives the appropriate references. The remaining targets where our HARPS observations could not detect a radial velocity signal are referred to here as null results. There are 12 such targets, which we model to derive upper limits on planetary mass, assuming the planet scenario. Some targets had archival CORALIE data \citep{Segransan2010} which we include in the fit where possible. Figure \ref{fig:nep_desert_all} shows the upper limits alongside the confirmed planetary sample.

To model the joint photometry and HARPS datasets we use the {\tt exoplanet} \citep{exoplanet:joss} Python code framework, which also makes use of {\tt starry} \citep{starryLuger2019} and {\tt PYMC3} \citep{SalvatierPymc3}. For the photometry, we utilise the available {\tess} light curve of each TOI, accounting for the exposure time of the data by oversampling the photometric model. Before fitting the model, the flux of the photometry is normalised to zero by dividing the individual light curves by the median of their out-of-transit points and subtracting unity. The model utilises a limb darkened transit model following the \citet{kipping2013} quadratic limb darkening parameterisation, and a Keplerian orbit model. The key model parameters are the stellar radius $R_{\star}$ in solar radii, the stellar mass $M_{\star}$ in solar masses, the orbital period $P_{\rm{orb}}$ in days, the planetary radii $R_{\rm{p}}$, the epoch $t_0$, the impact parameter $b$, the eccentricity $e$, and the argument of periastron $\omega$. To derive upper limits we set the eccentricity to zero in all cases. Individual cases were fit with linear or quadratic drift in the radial velocities, and/or a Gaussian process (GP) to model variability in the photometry. GPs for the photometry were enacted with the SHOTerm kernel in \textsc{exoplanet}\footnote{\url{https://celerite2.readthedocs.io/en/latest/api/python/\#celerite2.terms.SHOTerm}}, representing a stochastically-driven, damped harmonic oscillator. Details of fitting choices for individual systems are in Table \ref{tab:models}. No GP was used for any radial velocities. 

To fit the model, we use the \verb|PyMC| sampler, which draws samples from the posterior using a variant of Hamiltonian Monte Carlo, the No-U-Turn Sampler (NUTS). We allow for 1500 burn-in samples which are discarded, followed by 5000 steps for each of 12 chains, leading to 60000 samples for each target. Upper limits are given as the 95\% confidence value for the planetary mass from this posterior distribution.

Two targets, TOI-855 and TOI-4461, showed a marginally significant radial velocity amplitude ($2.8\sigma$ and $2.4\sigma$ respectively). We do not claim these targets as confirmed planets here, as this would require a focused study beyond the scope of the paper, but present the fitted values for the planetary mass in Table \ref{tab:planet_prop} and include them in our sample. TOI-855 additionally shows a quadratic drift in the radial velocities with coefficients $-0.001149\pm0.000009$ms$^{-1}$d$^{-2}$ (quad) and $3.338\pm0.004$ms$^{-1}$d$^{-1}$ (linear). Our radial velocities in this case span an approx. 4 year period, implying an additional radial velocity signal with amplitude of order kms$^{-1}$ on a period of years to decades.



\begin{table*}
    \caption{Model setup for joint fit targets in this paper. `Quad' is short for quadratic. `LC' is short for lightcurve.}

    \centering
    
    \begin{tabular}{lllllllr}
        \hline
        \hline
        Star & \textit{TESS} sectors & \textit{TESS} Cadence (min) & N. \textit{HARPS} & N. \textit{CORALIE} & GP (Photometry)  & RV drift & Notes \\
        \hline
       TOI-355  & 2,3,29,30,69 & 30,30,2,2,2 & 13 & 7 &  N & Quad & Drift 81.6ms$^{-1}$d$^{-1}$ \\
       TOI-422  & 5,32 & 2,2 & 6 & -- & Yes  & N &\\
       TOI-499  & 7,8,34,35,61,62 & 2 (all) & 12 & 2 &  N & N & \\
       TOI-728  & 8,35 & 2,2 & 5 & 2 & N  & N & \\
       TOI-851  & 3,30 & 30,2 & 7 & -- &  N & Linear & Drift 3.4ms$^{-1}$d$^{-1}$\\
       TOI-855  & 3,30,42,43 & 30,10,2,2 & 39 & 7 & N  & Quad & Marginal planet detection\\
       TOI-855  & &  & &  &   &  & Stellar companion from drift\\
       TOI-1975  & 10,11,37,64 & 30,30,2,2 & 15 & -- & N  & N & \\
       TOI-2227 & 1,27,28,67,68 & 2 (all) & 7 & -- & N  & N & \\
       TOI-2358  & 10,64 & 30,2 & 17 & -- & N  & N & \\
       TOI-2365  & 6,7,8,33,35,61,62 & 30,30,30,10,10,2,2 & 10 & -- & N  & N & \\
       TOI-4461  & 26,40,53 & 2,10,10 & 23 & -- & N  & N & Marginal planet detection \\
       TOI-4537  & 42,70 & 2,2 & 7 & -- & Yes  & N & LC variability in phase with transit\\      
        \hline
    \end{tabular}
    \label{tab:models}
\end{table*}

\bsp	
\label{lastpage}
\end{document}